\documentclass[11pt]{article}

\usepackage{graphicx,amsmath,amsfonts,amssymb,bm,hyperref,url,breakurl,epsfig,epsf,color,fullpage,MnSymbol,mathbbol,fmtcount,algorithmic,algorithm,semtrans,cite,caption,subcaption,multirow}
  
  \usepackage{mathtools}

\usepackage{titlesec}

\usepackage{tikz}
\usepackage{pgfplots}
\usetikzlibrary{pgfplots.groupplots}

\titleformat{\paragraph}
{\normalfont\normalsize\bfseries}{\theparagraph}{1em}{}
\titlespacing*{\paragraph}
{0pt}{3.25ex plus 1ex minus .2ex}{1.5ex plus .2ex}

\usepackage{movie15}

\usepackage{cite}
\usepackage{caption}
\usepackage[bottom,hang,flushmargin]{footmisc} 

\usepackage{hyperref}
\definecolor{darkred}{RGB}{150,0,0}
\definecolor{darkgreen}{RGB}{0,150,0}
\definecolor{darkblue}{RGB}{0,0,200}
\hypersetup{colorlinks=true, linkcolor=darkred, citecolor=darkgreen, urlcolor=darkblue}

\newtheorem{theorem}{Theorem}[section]
\newtheorem{lemma}[theorem]{Lemma}
\newtheorem{corollary}[theorem]{Corollary}

\newtheorem{definition}[theorem]{Definition}
\newtheorem{condition}[theorem]{Condition}

\newcommand{\opnorm}[1]{\left\|#1\right\|}

\newcommand{\twonorm}[1]{\left\|#1\right\|}

\newcommand{\abs}[1]{\left|#1\right|}
\newcommand{\avg}[1]{\left< #1 \right>}

\renewcommand{\d}{\mathrm{d}}

\newcommand{\x}{\vct{x}}

\definecolor{emmanuel}{RGB}{255,127,0}

\newcommand{\R}{\mathbb{R}}
\newcommand{\C}{\mathbb{C}}

\newcommand{\<}{\langle}
\renewcommand{\>}{\rangle}

\renewcommand{\P}{\operatorname{\mathbb{P}}}
\newcommand{\E}{\operatorname{\mathbb{E}}}

\newcommand{\vct}[1]{\bm{#1}}
\newcommand{\mtx}[1]{\bm{#1}}

\newcommand{\Real}{\operatorname{Re}}
\newcommand{\Imag}{\operatorname{Im}}

\numberwithin{equation}{section} 

\def \endprf{\hfill {\vrule height6pt width6pt depth0pt}\medskip}
\newenvironment{proof}{\noindent {\bf Proof} }{\endprf\par}

\title{Phase Retrieval via Wirtinger Flow: Theory and Algorithms}
\author{Emmanuel J. Cand\`{e}s\thanks{Departments of Mathematics and
    of Statistics, Stanford University, Stanford CA} \quad  Xiaodong
  Li\thanks{Department of Statistics, The Wharton School, University
    of Pennsylvania, Philadelphia, PA} \quad Mahdi
  Soltanolkotabi\thanks{Ming Hsieh Department of Electrical Engineering, University of Southern California, Los Angeles, CA} }

\date{July 3, 2014; Revised January 2015}

\begin{document}
\maketitle

\begin{abstract}
  We study the problem of recovering the phase from magnitude
  measurements; specifically, we wish to reconstruct a complex-valued
  signal $\vct{x}\in\C^n$ about which we have phaseless samples of the
  form $y_r = \abs{\langle \vct{a}_r,\vct{x} \rangle}^2$, $r = 1,
  \ldots, m$ (knowledge of the phase of these samples would yield a
  linear system). This paper develops a non-convex formulation of the
  phase retrieval problem as well as a concrete solution algorithm.
  In a nutshell, this algorithm starts with a careful initialization
  obtained by means of a spectral method, and then refines this
  initial estimate by iteratively applying novel update rules, which
  have low computational complexity, much like in a gradient descent
  scheme.  The main contribution is that this algorithm is shown to
  rigorously allow the exact retrieval of phase information from a
  nearly minimal number of random measurements. Indeed, the sequence
  of successive iterates provably converges to the solution at a
  geometric rate so that the proposed scheme is efficient both in
  terms of computational and data resources. In theory, a variation on
  this scheme leads to a near-linear time algorithm for a physically
  realizable model based on coded diffraction patterns. We illustrate
  the effectiveness of our methods with various experiments on image
  data. Underlying our analysis are insights for the analysis of
  non-convex optimization schemes that may have implications for
  computational problems beyond phase retrieval.
\end{abstract}

\section{Introduction}

We are interested in solving quadratic equations of the form
\begin{align}
\label{quadeq}
y_r=\abs{\<\vct{a}_r, \vct{z}\>}^2,\quad r=1,2,\ldots,m,
\end{align}
where $\vct{z}\in\C^n$ is the decision variable, $\vct{a}_r\in\C^n$
are known sampling vectors, and $y_r\in\R$ are observed
measurements. This problem is a general instance of a nonconvex
quadratic program (QP). Nonconvex QPs have been observed to occur
frequently in science and engineering and, consequently, their study
is of importance.  For example, a class of combinatorial optimization
problems with Boolean decision variables may be cast as QPs
\cite[Section 4.3.1]{nemirovski2001lectures}. Focusing on the
literature on physical sciences, the problem \eqref{quadeq} is
generally referred to as the phase retrieval problem. To understand
this connection, recall that most detectors can only record the
intensity of the light field and not its phase. Thus, when a small
object is illuminated by a quasi-monochromatic wave, detectors measure
the magnitude of the diffracted light. In the far field, the
diffraction pattern happens to be the Fourier transform of the object
of interest---this is called Fraunhofer diffraction---so that in
discrete space, \eqref{quadeq} models the data aquisition mechanism in
a coherent diffraction imaging setup; one can identify $\vct{z}$ with
the object of interest, $\vct{a}_r$ with complex sinusoids, and $y_r$
with the recorded data. Hence, we can think of \eqref{quadeq} as a
{\em generalized phase retrieval} problem. As is well known, the phase
retrieval problem arises in many areas of science and engineering such
as X-ray crystallography \cite{harrison1993phase, millane1990phase},
microscopy \cite{miao2008extending}, astronomy \cite{fienup1987phase},
diffraction and array imaging \cite{bunk2007diffractive,
  chai2011array}, and optics \cite{walther1963question}. Other fields
of application include acoustics \cite{balan2006signal,
  balan2010signal}, blind channel estimation in wireless
communications \cite{ahmed2012blind, ranieri2013phase}, interferometry
\cite{demanet2013convex}, quantum mechanics \cite{corbett2006pauli,
  reichenbach1965philosophic} and quantum information
\cite{heinosaari2013quantum}.  We refer the reader to the tutorial
paper \cite{shechtman2014phase} for a recent review of the theory and
practice of phase retrieval.

Because of the practical significance of the phase retrieval problem
in imaging science, the community has developed methods for recovering
a signal $\x \in \C^n$ from data of the form $y_r = |\<\vct{a}_r,
\vct{x}\>|^2$ in the special case where one samples the (square)
modulus of its Fourier transform. In this setup, the most widely used
method is perhaps the error reduction algorithm and its
generalizations, which were derived from the pioneering research of
Gerchberg and Saxton \cite{gerchberg1972practical} and Fienup
\cite{fienup1978reconstruction, fienup1982phase}.  The
Gerchberg-Saxton algorithm starts from a random initial estimate and
proceeds by iteratively applying a pair of `projections': at each
iteration, the current guess is projected in data space so that the
magnitude of its frequency spectrum matches the observations; the
signal is then projected in signal space to conform to some a-priori
knowledge about its structure. In a typical instance, our knowledge
may be that the signal is real-valued, nonnegative and spatially
limited.  First, while error reduction methods often work well in
practice, the algorithms seem to rely heavily on a priori information
about the signals, see \cite{hayes1982reconstruction, sanz1984note,
  fienup1982fine, fienup1983comments}. Second, since these algorithms
can be cast as alternating projections onto nonconvex sets
\cite{bauschke2003hybrid} (the constraint in Fourier space is not
convex), fundamental mathematical questions concerning their
convergence remain, for the most part, unresolved; we refer to Section
\ref{compnoncvx} for further discussion.

  On the theoretical side, several combinatorial optimization
  problems---optimization programs with discrete design variables
  which take on integer or Boolean values---can be cast as solving
  quadratic equations or as minimizing a linear objective subject to
  quadratic inequalities. In their most general form these problems are known to be
  notoriously difficult (NP-hard) \cite[Section
  4.3]{nemirovski2001lectures}. Nevertheless, many heuristics have
  been developed for addressing such problems.\footnote{For a partial
    review of some of these heuristics as well as some recent
    theoretical advances in related problems we refer to our companion
    paper \cite[Section 1.6]{PRCDP} and references therein
    \cite{cahillphase, ohlsson2011compressive,
      oymak2012simultaneously, bandeira2013saving,
      waldspurger2012phase,
      jaganathan2012robust,gross2013partial,gross2014improved}.}  One
  popular heuristic is based on a class of convex relaxations known as
  Shor's relaxations \cite[Section 4.3.1]{nemirovski2001lectures}
  which can be solved using tractable semi-definite programming
  (SDP). For certain random models, some recent SDP relaxations such
  as PhaseLift \cite{candes2013phase} are known to provide exact
  solutions (up to global phase) to the generalized phase retrieval
  problem using a near minimal number of sampling vectors
  \cite{candes2012phaselift,PRCDP}. While in principle SDP based
  relaxations offer tractable solutions, they become computationally
  prohibitive as the dimension of the signal increases. Indeed, for a
  large number of unknowns in the tens of thousands, say, the memory
  requirements are far out of reach of desktop computers so that these
  SDP relaxations are de facto impractical.

\section{Algorithm: Wirtinger Flow}

This paper introduces an approach to phase retrieval based on
non-convex optimization as well as a solution algorithm, which has two
components: (1) a careful initialization obtained by means of a
spectral method, and (2) a series of updates refining this initial
estimate by iteratively applying a novel update rule, much like in a
gradient descent scheme. We refer to the combination of these two
steps, introduced in reverse order below, as the {\em Wirtinger flow}
(WF) algorithm.

\subsection{Minimization of a non-convex objective}

Let $\ell(x,y)$ be a loss function measuring the misfit between both
its scalar arguments. If the loss function is non-negative and
vanishes only when $x = y$, then a solution to the generalized phase
retrieval problem \eqref{quadeq} is any solution to
\begin{equation}
\label{noncvx}
\text{minimize} \quad f(\vct{z}):=\frac{1}{2m}\sum_{r=1}^m
\ell\bigl(y_r, \abs{\vct{a}_r^*\vct{z}}^2\bigr), \quad \vct{z} \in \C^n.  
\end{equation}
Although one could study many loss functions, we shall focus in this
paper on the simple quadratic loss $\ell(x,y)=(x-y)^2$. Admittedly,
the formulation \eqref{noncvx} does not make the problem any easier
since the function $f$ is not convex. Minimizing non-convex
objectives, which may have very many stationary points, is known to be
NP-hard in general.  In fact, even establishing convergence to a local
minimum or stationary point can be quite challenging, please see
\cite{murty1987some} for an example where convergence to a local
minimum of a degree-four polynomial is known to be
NP-hard.\footnote{Observe that if all the sampling vectors are real
  valued, our objective is also a degree-four polynomial.} As a side
remark, deciding whether a stationary point of a polynomial of degree
four is a local minimizer is already known to be NP-hard.

Our approach to \eqref{noncvx} is simply stated: start with an
initialization $\vct{z}_0$, and for $\tau = 0, 1, 2, \ldots$,
inductively define
\begin{equation}
\label{graddescent}
\vct{z}_{\tau+1}=\vct{z}_\tau-\frac{\mu_{\tau+1}}{\twonorm{\vct{z}_0}^2}\left(\frac{1}{m}\sum_{r=1}^m\left(\abs{\vct{a}_r^*\vct{z}}^2-y_r\right)(\vct{a}_r\vct{a}_r^*)\vct{z}\right):=\vct{z}_\tau-\frac{\mu_{\tau+1}}{\twonorm{\vct{z}_0}^2}\nabla
f(\vct{z}_\tau). 
\end{equation}
If the decision variable $\vct{z}$ and the sampling vectors were all
real valued, the term between parentheses would be the gradient of
$f$ divided by two, as our notation suggests. However, since $f(\vct{z})$ is a
mapping from $\C^n$ to $\R$, it is not holomorphic and hence not
complex-differentiable. However, this term can still be viewed as a
gradient based on Wirtinger derivatives reviewed in Section
\ref{wirt}. Hence, \eqref{graddescent} is a form of steepest descent
and the parameter $\mu_{\tau+1}$ can be interpreted as a step size
(note nonetheless that the effective step size is also inversely
proportional to the magnitude of the initial guess).

\subsection{Initialization via a spectral method}
\label{Initschemes}

Our main result states that for a certain random model, if the
initialization $\vct{z}_0$ is sufficiently accurate, then the sequence
$\{\vct{z}_{\tau}\}$ will converge toward a solution to the
generalized phase problem \eqref{quadeq}.  In this paper, we propose
computing the initial guess $\vct{z}_0$ via a spectral method,
detailed in Algorithm \ref{QIINIT}. In words, $\vct{z}_0$ is the
leading eigenvector of the positive semidefinite Hermitian matrix
$\sum_r y_r \vct{a}_r \vct{a}_r^*$ constructed from the knowledge of
the sampling vectors and observations. (As usual, $\vct{a}_r^*$ is the
adjoint of $\vct{a}_r$.) Letting $\mtx{A}$ be the $m \times n$ matrix
whose $r$th row is $\vct{a}_r^*$ so that with obvious notation
$\vct{y} = |\mtx{A} \vct{x}|^2$, $\vct{z}_0$ is the leading
eigenvector of $\mtx{A}^* \, \operatorname{diag}\{\vct{y}\}\, \mtx{A}$
and can be computed via the power method by repeatedly applying
$\mtx{A}$, entrywise multiplication by $\vct{y}$ and $\mtx{A}^*$. In
the theoretical framework we study below, a constant number of power
iterations would give machine accuracy because of an eigenvalue gap
between the top two eigenvalues, please see Appendix \ref{comp} for
additional information.
\begin{algorithm}[h]
  \caption{Wirtinger Flow: Initialization} 
\begin{algorithmic}
  \REQUIRE{Observations $\{y_r\} \in \R^m$.}
  \STATE Set 
\[ 
\lambda^2 = n \, \frac{\sum_{r}
    y_r}{\sum_r \twonorm{\vct{a}_r}^2}.
\]
\STATE Set $\vct{z}_0$, normalized to $\|\vct{z}_0\| = \lambda$, to be
the eigenvector corresponding to the largest eigenvalue of
  \begin{align*}
  \mtx{Y}=\frac{1}{m}\sum_{r=1}^my_r\vct{a}_r\vct{a}_r^*.
  \end{align*}
  \ENSURE{Initial guess $\vct{z}_0$.}
\end{algorithmic}
\label{QIINIT}
\end{algorithm}

\subsection{Wirtinger flow as a stochastic gradient scheme}

We would like to motivate the Wirtinger flow algorithm and provide
some insight as to why we expect it to work in a model where the
sampling vectors are random.  First, we emphasize that our statements
in this section are heuristic in nature; as it will become clear in
the proof Section \ref{proofsec}, a correct mathematical formalization
of these ideas is far more complicated than our heuristic development
here may
suggest. 
Second, although our ideas are broadly applicable, it makes sense to
begin understanding the algorithm in a setting where everything is
real valued, and in which the vectors $a_r$ are
i.i.d.~$\mathcal{N}(\vct{0},\mtx{I})$. Also without any loss in generality and to simplify exposition in this section we shall assume $\twonorm{\vct{x}}=1$.

Let $\vct{x}$ be a solution to \eqref{quadeq} so that $y_r = |\<
\vct{a}_r, \vct{x}\>|^2$, and consider the initialization step first.
In the Gaussian model, a simple moment calculation gives
\[
\E \left[\frac{1}{m}\sum_{r=1}^my_r\vct{a}_r\vct{a}_r^*\right] =
\mtx{I} + 2 \vct{x}\vct{x}^*. 
\]
By the strong law of large numbers, the matrix $\mtx{Y}$ in Algorithm
\ref{QIINIT} is equal to the right-hand side in the limit of large
samples. Since any leading eigenvector of $\mtx{I} + 2
\vct{x}\vct{x}^*$ is of the form $\lambda \vct{x}$ for some scalar
$\lambda \in \R$, we see that the intialization step would recover
$\vct{x}$ perfectly, up to a global sign or phase factor, had we
infinitely many samples. Indeed, the chosen normalization would
guarantee that the recovered signal is of the form $\pm \vct{x}$.  As
an aside, we would like to note that the top two eigenvalues of
$\mtx{I} + 2 \vct{x}\vct{x}^*$ are well separated unless $\|\vct{x}\|$
is very small, and that their ratio is equal to $1 + 2
\|\vct{x}\|^2$.  Now with a finite amount of data, the leading
eigenvector of $\mtx{Y}$ will of course not be perfectly correlated
with $\vct{x}$ but we hope that it is sufficiently correlated to point
us in the right direction.

We now turn our attention to the gradient-update \eqref{graddescent}
and define
\begin{align*}
F(\vct{z})=\vct{z}^*(\mtx{I}-\vct{x}\vct{x}^*)\vct{z}+\frac{3}{4}\left(\twonorm{\vct{z}}^2-1\right)^2,
\end{align*}

where here and below, $\vct{x}$ is once again our planted
solution. The first term ensures that the direction of $\vct{z}$
matches the direction of $\vct{x}$ and the second term penalizes the
deviation of the Euclidean norm of $\vct{z}$ from that of
$\vct{x}$. Obviously, the minimizers of this function are $\pm
\vct{x}$. Now consider the gradient scheme
\begin{align}
\label{iterbar}
\vct{z}_{\tau+1}=\vct{z}_\tau-\frac{\mu_{\tau+1}}{\twonorm{\vct{z}_0}^2}\nabla
F(\vct{z}_\tau).
\end{align}
In Section \ref{QIINITsec}, we show that if $\text{min }
\twonorm{\vct{z}_0 \pm \vct{x}}\le 1/8\twonorm{\vct{x}}$, then
$\{\vct{z}_\tau\}$ converges to $\vct{x}$ up to a global sign.
However, this is all ideal as we would need knowledge of $\vct{x}$
itself to compute the gradient of $F$; we simply cannot run this
algorithm.

Consider now the WF update and assume for a moment that $\vct{z}_\tau$
is fixed and independent of the sampling vectors. We are well aware
that this is a false assumption but nevertheless wish to explore some
of its consequences.  In the Gaussian model, if $\vct{z}$ is
independent of the sampling vectors, then a modification of Lemma \ref{expGrad} for real-valued $\vct{z}$ shows
that $\E[\nabla f(\vct{z})]=\nabla F(\vct{z})$ and, therefore,
\begin{align*}
  \E[\vct{z}_{\tau+1}]=\E[\vct{z}_\tau]-\frac{\mu_{\tau+1}}{\twonorm{\vct{z}_0}^2}\E[\nabla
  f(\vct{z}_\tau)]\quad\Rightarrow\quad\E[\vct{z}_{\tau+1}]=\vct{z}_\tau-\frac{\mu_{\tau+1}}{\twonorm{\vct{z}_0}^2}\nabla
  F(\vct{z}_\tau).
\end{align*}
Hence, the average WF update is the same as that in \eqref{iterbar} so
that we can interpret the Wirtinger flow algorithm as a stochastic
gradient scheme in which we only get to observe an unbiased estimate
$\nabla f(\vct{z})$ of the ``true'' gradient $\nabla F(\vct{z})$.

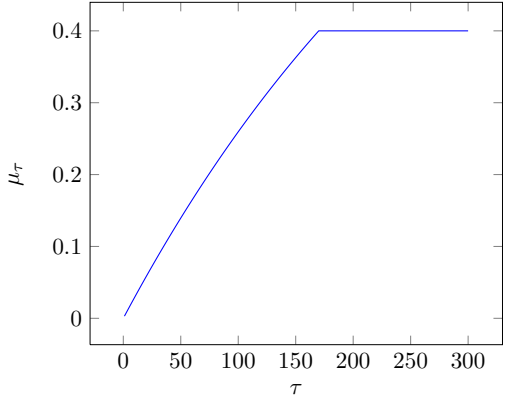
\begin{figure}[h]
        \centering
\begin{tikzpicture}[scale=0.8] 
\begin{groupplot}[group style={group size=1 by 1, xlabels at=edge bottom, ylabels at=edge left,xticklabels at=edge bottom},xlabel=$\tau$,
        ylabel=$\mu_\tau$]
 \nextgroupplot[]
        \addplot +[mark=solid,blue,line width=0.5pt] table[x index=0,y index=1]{./lparam};
\end{groupplot}
\end{tikzpicture}
\caption{Learning parameter $\mu_\tau$ from \eqref{learnparam} as a
  function of the iteration count $\tau$; here, $\tau_0 \approx 330$
  and $\mu_{\text{max}}=0.4$.}
 \label{learnparamfig}
\end{figure}

Regarding WF as a stochastic gradient scheme helps us in choosing the
learning parameter or step size $\mu_\tau$. Lemma \ref{concenGrad}
asserts that
\begin{equation}
\label{vargrad}
\twonorm{ \nabla f(\vct{z}) - \nabla F(\vct{z})}^2 \le
\twonorm{\vct{x}}^2 \cdot \min  \,\,\, \twonorm{\vct{z} \pm \vct{x}}
\end{equation}
holds with high probability.  Looking at the right-hand side, this
says that the uncertainty about the gradient estimate depends on how
far we are from the actual solution $\vct{x}$. The further away, the
larger the uncertainty or the noisier the estimate.  This suggests
that in the early iterations we should use a small learning parameter
as the noise is large since we are not yet close to the
solution. However, as the iterations count increases and we make
progress, the size of the noise also decreases and we can pick larger
values for the learning parameter. This heuristic together with
experimentation lead us to consider
\begin{align}
\label{learnparam}
\mu_\tau=\min(1-e^{-\tau/\tau_0},\mu_{\text{max}})
\end{align}
shown in Figure \ref{learnparamfig}. Values of $\tau_0$ around $330$
and of $\mu_{\text{max}}$ around $0.4$ worked well in our simulations.
This makes sure that $\mu_\tau$ is rather small at the beginning
(e.g.~$\mu_1\approx 0.003$ but quickly increases and reaches a maximum
value of about $0.4$ after 200 iterations or so.
 
\section{Main Results}

\subsection{Exact phase retrieval via Wirtinger flow}

Our main result establishes the correctness of the Wirtinger flow algorithm in the
Gaussian model defined below. Later in Section \ref{sec:Fourier}, we
shall also develop exact recovery results for a physically inspired
diffraction model.
\begin{definition}
  \label{def:gaussian}
  We say that the sampling vectors follow the Gaussian model if
  $\vct{a}_r \in \C^n \stackrel{\mathsf{i.i.d.}}{\sim}
  \mathcal{N}(\vct{0},\mtx{I}/2) + i
  \mathcal{N}(\vct{0},\mtx{I}/2)$. In the real-valued case, they are
  i.i.d.~$ \mathcal{N}(\vct{0},\mtx{I})$.
\end{definition}
We also need to define the distance to the solution set. 
\begin{definition}
  Let $\vct{x}\in\C^n$ be any solution to the quadratic system
  \eqref{quadeq} (the signal we wish to recover). For each
  $\vct{z}\in\C^n$, define
\[
\emph{dist}(\vct{z},\vct{x}) = \min_{\phi\in[0,2\pi]}
\,\,\, \twonorm{\vct{z}-e^{i\phi}\vct{x}}.
\]
\end{definition}

\begin{theorem}
\label{gauss1} 
Let $\vct{x}$ be an arbitrary vector in $\C^n$ and $\vct{y} = |\mtx{A}
\vct{x}|^2 \in \R^m$ be $m$ quadratic samples with $m\ge c_0 \cdot
n\log n$, where $c_0$ is a sufficiently large numerical constant. Then
the Wirtinger flow initial estimate $\vct{z}_0$ normalized to
have squared Euclidean norm equal to $m^{-1} \sum_r y_r$,\footnote{The
  same results holds with the intialization from Algorithm
  \ref{QIINIT} because $\sum_r \|a_r\|^2 \approx m \cdot n$ with a
  standard deviation of about the square root of this quantity.} obeys
\begin{align}
\label{gauss1init}
\text{\em dist}(\vct{z}_0,\vct{x}) \le \frac{1}{8}\twonorm{\vct{x}}
\end{align}
with probability at least $1-10e^{-\gamma n}-{8}/{n^2}$ ($\gamma$
is a fixed positive numerical constant). Further, take a constant
learning parameter sequence, $\mu_\tau= \mu$ for all $\tau = 1, 2,
\ldots$ and assume $\mu \le c_1/n$ for some fixed numerical constant
$c_1$.  Then there is an event of probability at least
$1-13e^{-\gamma n}-me^{-1.5m}-{8}/{n^2}$, such that on this event,
starting from any initial solution $\vct{z}_0$ obeying
\eqref{gauss1init}, we have
\begin{equation*}
  \text{\em dist}(\vct{z}_\tau,\vct{x})\le
  \frac{1}{8}\left(1-\frac{\mu}{4}\right)^{\tau/2} \cdot
  \twonorm{\vct{x}}.
\end{equation*} 
\end{theorem}
  
Clearly, one would need $2n$ quadratic measurements to have any hope
of recovering $x \in \C^n$. It is also known that in our sampling
model, the mapping $\vct{z} \mapsto |\mtx{A} \vct{z}|^2$ is injective
for $m \ge 4n$ \cite{balan2006signal} and that this property holds for
generic sampling vectors \cite{conca2013algebraic}.\footnote{It is not
  within the scope of this paper to explain the meaning of generic
  vectors and, instead, refer the interested reader to
  \cite{conca2013algebraic}.} Hence, the Wirtinger flow algorithm
loses at most a logarithmic factor in the {\em sampling
  complexity}. In comparison, the SDP relaxation only needs a sampling
complexity proportional to $n$ (no logarithmic factor)
\cite{candes2012solving}, and it is an open question whether Theorem
\ref{gauss1} holds in this regime.

Setting $\mu=c_1/n$ yields $\epsilon$ accuracy in a relative sense,
namely, $\operatorname{dist}(\vct{z},\vct{x})\le
\epsilon\twonorm{\vct{x}}$, in $\mathcal{O}(n\, \log 1/\epsilon)$
iterations. The computational work at each iteration is dominated by
two matrix-vector products of the form $\mtx{A} \vct{z}$ and
$\mtx{A}^* \vct{v}$, see Appendix \ref{comp}.  It follows that the
overall computational complexity of the WF algorithm is
$\mathcal{O}(mn^2\log{1}/{\epsilon})$. Later in the paper, we will
exhibit a modification to the WF algorithm of mere theoretical
interest, which also yields exact recovery under the same sampling
complexity and an $\mathcal{O}(mn\log{1}/{\epsilon})$ computational
complexity; that is to say, the computational workload is now just
{\em linear} in the problem size.

\subsection{Comparison with other non-convex schemes}
\label{compnoncvx}

We now pause to comment on a few other non-convex schemes in the
literature. Other comparisons may be found in our companion paper
\cite{PRCDP}.

Earlier, we discussed the Gerchberg-Saxton and Fienup
algorithms. These formulations assume that $\mtx{A}$ is a Fourier
transform and can be described as follows: suppose $\vct{z}_\tau$ is
the current guess, then one computes the image of $\vct{z}_\tau$
through $\mtx{A}$ and adjust its modulus so that it matches that of
the observed data vector: with obvious notation, 
\begin{equation}
  \label{eq:GS0}
\hat{\vct{v}}_{\tau+1} = \vct{b} \odot \frac{\mtx{A}
  \vct{z}_\tau}{|\mtx{A} \vct{z}_\tau|},
\end{equation}
where $\odot$ is elementwise multiplication, and $\vct{b} = |\mtx{A}
\vct{x}|$ so that $b_r^2 = y_r$ for all $r = 1, \ldots, m$.  Then
\begin{equation}
  \label{eq:GS1}
  \vct{v}_{\tau+1} = \text{arg min}_{\vct{v} \in \C^n} 
  \,\,\, \|\hat{\vct{v}}_{\tau+1} - \mtx{A} \vct{v}\|.
\end{equation}
(In the case of Fourier data, the step \eqref{eq:GS0}--\eqref{eq:GS1}
essentially adjusts the modulus of the Fourier transform of the
current guess so that it fits the measured data.) Finally, if we know
that the solution belongs to a convex set $\mathcal{C}$ (as in the
case where the signal is known to be real-valued, possibly
non-negative and of finite support), then the next iterate is
\begin{equation}
  \label{eq:GS2}
\vct{z}_{\tau+1} = P_{\mathcal{C}}(\vct{v}_{\tau+ 1}), 
\end{equation}
where $P_{\mathcal{C}}$ is the projection onto the convex set
$\mathcal{C}$. If no such information is available, then
$\vct{z}_{\tau+1} = \vct{v}_{\tau+1}$. The first step \eqref{eq:GS1}
is not a projection onto a convex set and, therefore, it is in general
completely unclear whether the Gerchberg-Saxton algorithm actually
converges. (And if it were to converge, at what speed?) It is also
unclear how the procedure should be initialized to yield accurate
final estimates. This is in contrast to the Wirtinger flow algorithm, which in the
Gaussian model is shown to exhibit geometric convergence to the
solution to the phase retrieval problem. Another benefit is that the
Wirtinger flow algorithm does not require solving a least-squares problem
\eqref{eq:GS1} at each iteration; each step enjoys a reduced
computational complexity.

A recent contribution related to ours is the interesting paper
\cite{netrapalli2013phase}, which proposes an alternating minimization
scheme named AltMinPhase for the general phase retrieval
problem. AltMinPhase is inspired by the Gerchberg-Saxton update
\eqref{eq:GS0}--\eqref{eq:GS1} as well as other established
alternating projection heuristics \cite{fienup1982phase,
  yang1994gerchberg, marchesini2007invited, marchesini2007phase,
  miao1999extending, miao2002high}. We describe the algorithm in the
setup of Theorem \ref{gauss1} for which \cite{netrapalli2013phase}
gives theoretical guarantees. To begin with, AltMinPhase partitions
the sampling vectors $\vct{a}_r$ (the rows of the matrix $\mtx{A}$)
and corresponding observations $y_r$ into $B+1$ disjoint blocks
$(\vct{y}^{(0)},\mtx{A}^{(0)})$, $(\vct{y}^{(1)},\mtx{A}^{(1)})$,
$\ldots$, $(\vct{y}^{(B)},\mtx{A}^{(B)})$ of roughly equal
size. Hence, distinct blocks are stochastically independent from each
other. The first block $(\vct{y}^{(0)},\mtx{A}^{(0)})$ is used to
compute an initial estimate $\vct{z}_0$.  After initialization,
AltMinPhase goes through a series of iterations of the form
\eqref{eq:GS0}--\eqref{eq:GS1}, however, with the key difference that
each iteration uses a fresh set of sampling vectors and observations:
in details,
 \begin{equation}
   \label{eq:ALTMIN}
   \vct{z}_{\tau+1} = \text{arg min}_{\vct{z} \in \C^n} 
  \,\,\, \|\hat{\vct{v}}_{\tau+1} - \mtx{A}^{(\tau+1)} \vct{z}\|, \qquad \hat{\vct{v}}_{\tau+1}  = \vct{b} \odot \frac{\mtx{A}^{(\tau+1)}
  \vct{z}_\tau}{|\mtx{A}^{(\tau+1)} \vct{z}_\tau|}.
\end{equation}
As for the Gerchberg-Saxton algorithm, each iteration requires solving
a least-squares problem. Now assume a real-valued Gaussian model as
well as a real valued solution $x \in \R^n$. The main result in
\cite{netrapalli2013phase} states that if the first block
$(\vct{y}^{(0)},\mtx{A}^{(0)})$ contains at least $c \cdot n \log^3 n$
samples and each consecutive block contains at least $c \cdot n \log
n$ samples---$c$ here denotes a positive numerical constant whose
value may change at each occurence---then it is possible to initialize
the algorithm via data from the first block in such a way that each
consecutive iterate \eqref{eq:ALTMIN} decreases the error
$\|\vct{z}_\tau - \vct{x}\|$ by 50\%; naturally, all of this holds in
a probabilistic sense. Hence, one can get $\epsilon$ accuracy in the
sense introduced earlier from a total of $c \cdot n \log n \cdot
(\log^2n + \log {1}/{\epsilon})$ samples.  Whereas the Wirtinger flow algorithm
achieves arbitrary accuracy from just $c \cdot n \log n$ samples,
these theoretical results would require an infinite number of
samples. This is, however, not the main point.

The main point is that in practice, it is not realistic to imagine (1)
that we will divide the samples in distinct blocks (how many blocks
should we form a priori?  of which sizes?) and (2) that we will use
measured data only once. With respect to the latter, observe that the
Gerchberg-Saxton procedure \eqref{eq:GS0}--\eqref{eq:GS1} uses all the
samples at each iteration. This is the reason why AltMinPhase is of
little practical value, and of theoretical interest only. As a matter
of fact, its design and study seem merely to stem from analytical
considerations: since one uses an independent set of measurements at
each iteration, $A^{(\tau+1)}$ and $z_\tau$ are stochastically
independent, a fact which considerably simplifies the convergence
analysis.  In stark contrast, the WF iterate uses all the samples at
each iteration and thus introduces some dependencies, which makes for
some delicate analysis.  Overcoming these difficulties is crucial
because the community is preoccupied with convergence properties of
algorithms one actually runs, like Gerchberg-Saxton
\eqref{eq:GS0}--\eqref{eq:GS1}, or would actually want to run.
Interestingly, it may be possible to use some of the ideas developed
in this paper to develop a rigorous theory of convergence for
algorithms in the style of Gerchberg-Saxton and Fienup, please see
\cite{MSthesis}. 

In a recent paper \cite{Altglobal}, which appeared on the arXiv
preprint server as the final version of this paper was under
preparation, the authors explore necessary and sufficient conditions
for the global convergence of an alternative minimization scheme with
generic sampling vectors.  The issue is that we do not know when these
conditions hold. Further, even when the algorithm converges, it does
not come with an explicit convergence rate so that is is not known
whether the algorithm converges in polynomial time. As before, some of
our methods as well as those from our companion paper \cite{PRCDP} may
have some bearing upon the analysis of this algorithm. Similarly, another class of nonconvex algorithms that have recently
been proposed in the literature are iterative algorithms based on
Generalized Approximate Message Passing (GAMP), see
\cite{rangan2011generalized} and \cite{schniter2012compressive} as
well as \cite{donoho2009message, bayati2011dynamics, bayati2012lasso}
for some background literature on AMP. In \cite{schniter2012compressive}, the authors demonstrate a favorable
 runtime for an algorithm of this nature. However, this does not come
with any theoretical guarantees.

Moving away from the phase retrieval problem, we would like to mention
some very interesting work on the matrix completion problem using
non-convex schemes by Montanari and coauthors
\cite{keshavan2010matrix,keshavan2010matrixnoise,keshavan2012efficient},
see also \cite{jain2013low, hardt2013provable, lee2013near,
  agarwal2012fast,balzano2013local, mroueh2014robust,
  mroueh2013quantization}. Although the problems and models are quite
different, there are some general similarities between the algorithm
named OptSpace in \cite{keshavan2010matrix} and ours. Indeed, OptSpace
operates by computing an initial guess of the solution to a low-rank
matrix completion problem by means of a spectral method. It then sets
up a nonconvex problem, and proposes an iterative algorithm for
solving it. Under suitable assumptions, \cite{keshavan2010matrix}
demonstrates the correctness of this method in the sense that OptSpace
will eventually converge to a low-rank solution, although it is not
shown to converge in polynomial time.

\section{Numerical Experiments}
\label{numsec}

We present some numerical experiments to assess the empirical
performance of the Wirtinger flow algorithm. Here, we mostly consider a model of
coded diffraction patterns reviewed below.

\subsection{The coded diffraction model}
\label{defCDP}

We consider an acquisition model, where we collect data of the form  
\begin{equation}
\label{mainCDPmodel}
  y_r = \left| \sum_{t = 0}^{n-1} x[t] \bar{d}_\ell(t) e^{-i2\pi k t/n}
  \right|^2, \quad r = (\ell, k), \quad \begin{array}{l} 0 \le k \le n-1\\
    1 \le \ell \le L
  \end{array};
  \end{equation}
  thus for a fixed $\ell$, we collect the magnitude of the diffraction
  pattern of the signal $\{x(t)\}$ modulated by the waveform/code
  $\{d_\ell(t)\}$. By varying $\ell$ and changing the modulation
  pattern $d_\ell$, we generate several views thereby creating a
  series of \emph{coded diffraction patterns} (CDPs).

  In this paper, we are mostly interested in the situation where the
  modulation patterns are random; in particular, we study a model in
  which the $d_\ell$'s are i.i.d.~distributed, each having
  i.i.d.~entries sampled from a distribution $d$. Our theoretical
  results from Section \ref{sec:Fourier} assume that $d$ is symmetric,
  obeys $|d| \le M$ as well as the moment conditions
\begin{align}
\label{momentcond}
\E d = 0, \quad \E d^2 = 0,\quad \E \abs{d}^4 =2{(\E \abs{d}^2)}^2.
\end{align}
A random variable obeying these assumptions is said to be {\em
  admissible}.  Since $d$ is complex valued we can have $\E d^2 = 0$
while $d \neq 0$. An example of an admissible random variable is $d =
b_1 b_2$, where $b_1$ and $b_2$ are independent and distributed as
\begin{align}
\label{eq:octanary}
b_1=\begin{cases}
 +1&\text{with prob.~} {1}/{4}\\
-1&\text{with prob.~} {1}/{4}\\
-i&\text{with prob.~} {1}/{4}\\
+i&\text{with prob.~} {1}/{4}
\end{cases}
\quad \text{and} \quad b_2=\begin{cases}
  \sqrt{2}/2 & \text{with prob.~}  {4}/{5}\\
\sqrt{3} & \text{with prob.~}  {1}/{5}
\end{cases}.
\end{align}
We shall refer to this distribution as an {\em octanary} pattern since
$d$ can take on eight distinct values.  The condition $\E[d^2]=0$ is
here to avoid unnecessarily complicated calculations in our
theoretical analysis.  In particular, we can also work with a {\em
  ternary} pattern in which $d$ is distributed as
\begin{equation}
\label{eq:ternary}
d=\begin{cases}
+1 &\quad\text{with prob.~} {1}/{4}\\
0  &\quad\text{with prob.~} {1}/{2}\\
-1 &\quad\text{with prob.~} {1}/{4}
\end{cases}. 
\end{equation}
We emphasize that these models are physically realizable in optical applications specially those that arise in microscopy. However, we should note that phase retrieval has many different applications and in some cases other models may be more convenient. We refer to our companion paper \cite{PRCDP} Section 2.2 for a discussion of other practically relevant models.

\subsection{The Gaussian and coded diffraction models}

We begin by examining the performance of the Wirtinger flow algorithm for
recovering random signals $\vct{x}\in\C^n$ under the Gaussian and
coded diffraction models. We are interested in signals of two
different types:
\begin{itemize}
\item \emph{Random low-pass signals.} Here, $\x$ is given by 
  \begin{align*} {x}[t]=\sum_{k=-(M/2-1)}^{M/2} (X_k+iY_k) e^{2\pi i
      (k-1)(t-1)/n},
\end{align*}
with $M = n/8$ and $X_k$ and $Y_k$ are i.i.d.~$\mathcal{N}(0,1)$.
\item \emph{Random Gaussian signals.} In this model, $\x \in \C^n$ is
  a random complex Gaussian vector with i.i.d.~entries of the form
  $x[t] = X+iY$ with $X$ and $Y$ distributed as $\mathcal{N}(0,1)$;
  this can be expressed as
  \begin{align*} {x}[t]=\sum_{k=-(n/2-1)}^{n/2} (X_k+iY_k) e^{2\pi i
      (k-1)(t-1)/n},
\end{align*}
where $X_k$ and $Y_k$ are are i.i.d.~$\mathcal{N}(0,1/8)$ so that the
low-pass model is a `bandlimited' version of this high-pass random
model (variances are adjusted so that the expected signal power is the
same).
\end{itemize}
Below, we set $n=128$, and generate one signal of each type which will
be used in all the experiments.

The initialization step of the Wirtinger flow algorithm is run by applying $50$
iterations of the power method outlined in Algorithm \ref{pow} from
Appendix \ref{comp}. In the iteration \eqref{graddescent}, we use the
parameter value $\mu_\tau=\min(1-\exp(-\tau/\tau_0),0.2)$ where
$\tau_0 \approx 330$. We stop after $2,500$ iterations, and report the
empirical probability of success for the two different signal
models. The empirical probability of succcess is an average over $100$
trials, where in each instance, we generate new random sampling
vectors according to the Gaussian or CDP models. We declare a trial
successful if the relative error of the reconstruction
$\operatorname{dist}(\hat{\vct{x}},\vct{x})/\twonorm{\vct{x}}$ falls
below $10^{-5}$.

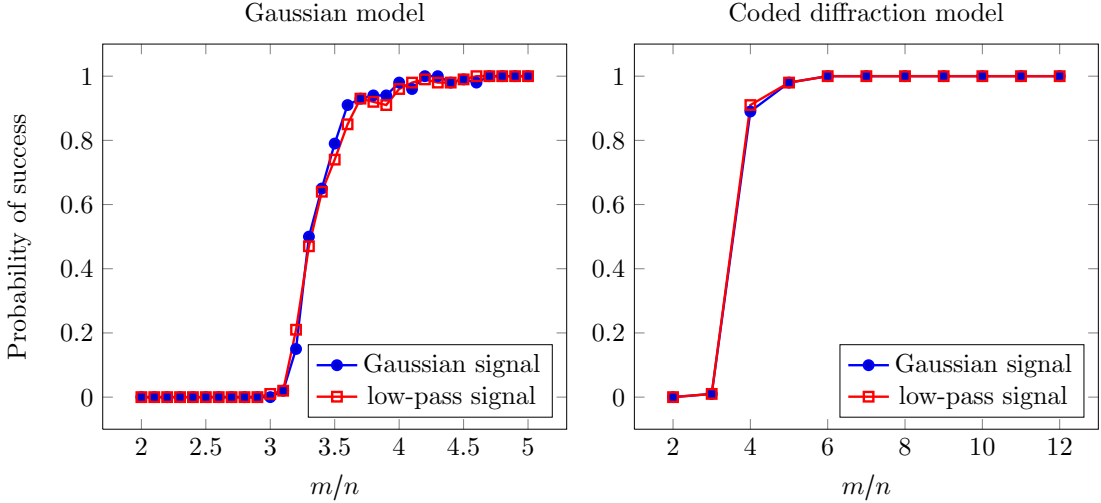
\begin{figure}[h]
        \centering
\begin{tikzpicture}[scale=0.9] 
\begin{groupplot}[group style={group size=2 by 1,horizontal sep=1cm,xlabels at=edge bottom, ylabels at=edge left,xticklabels at=edge bottom},xlabel=${m}/{n}$,
        ylabel=Probability of success,
        legend pos= south east]
 \nextgroupplot[title={Gaussian model}]
        
        \addplot +[mark=*,solid,blue,line width=1pt] table[x index=0,y index=1]{./probG};
        \addlegendentry{Gaussian signal}
        \addplot +[mark=square ,solid,red,line width=1pt] table[x index=0,y index=1]{./probGS};
        \addlegendentry{low-pass signal}

 \nextgroupplot[title={Coded diffraction model}]
        \addplot +[mark=*,solid,blue,line width=1pt] table[x index=0,y index=1]{./probCDP};\addlegendentry{Gaussian signal}
        \addplot +[mark=square ,solid,red,line width=1pt] table[x index=0,y index=1]{./probCDPS};\addlegendentry{low-pass signal}
 \end{groupplot}
\end{tikzpicture}
\caption{ Empirical probability of success based on $100$ random
  trials for different signal/measurement models and a varied number
  of measurements. The coded diffraction model uses octanary patterns;
  the number of patterns $L = m/n$ only takes on integral values.}
\label{PT1}
\end{figure}

Figure \ref{PT1} shows that around $4.5n$ Gaussian phaseless
measurements suffice for exact recovery with high probability via the
Wirtinger flow algorithm. We also see that about six octanary patterns are
sufficient.

\subsection{Performance on natural images}

We move on to testing the Wirtinger flow algorithm on various images of different
sizes; these are photographs of the Naqsh-e Jahan Square in the
central Iranian city of Esfahan, the Stanford main quad, and the Milky
Way galaxy. Since each photograph is in color, we run the WF algorithm
on each of the three RGB images that make up the photograph. Color
images are viewed as $n_1 \times n_2 \times 3$ arrays, where the first
two indices encode the pixel location, and the last the color band.

We generate $L=20$ random octanary patterns and gather the coded
diffraction patterns for each color band using these 20 samples.  As
before, we run 50 iterations of the power method as the initialization
step. The updates use the sequence
$\mu_\tau=\min(1-\exp({-\tau/\tau_0}),0.4)$ where $\tau_0 \approx 330$
as before. In all cases we run $300$ iterations and record the
relative recovery error as well as the running time. If $\vct{x}$ and
$\hat{\vct{x}}$ are the original and recovered images, the relative
error is equal to $\twonorm{\hat{\vct{x}}-\vct{x}}/\twonorm{\vct{x}}$,
where $\twonorm{\cdot}$ is the Euclidean norm $\twonorm{\vct{x}}^2 =
\sum_{i,j,k} |x(i,j,k)|^2$.  The computational time we report is the
the computational time averaged over the three RGB images. All
experiments were carried out on a MacBook Pro with a 2.4 GHz Intel
Core i7 Processor and 8 GB 1600 MHz DDR3 memory.

\begin{figure}
        \centering
         \begin{subfigure}[b]{\textwidth}
                \centering
                \includegraphics[width=0.9\textwidth]{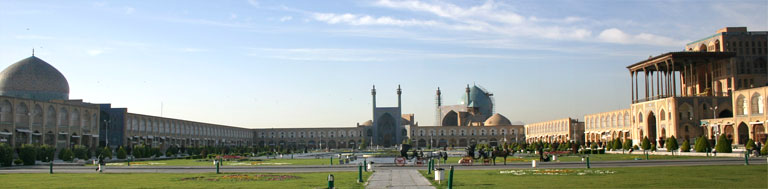}
                \begin{minipage}[t]{5in}\caption{Naqsh-e Jahan Square, Esfahan. Image size is
                  $189\times 768$ pixels; timing is 61.4 sec or about
                  21,200 FFT units. The relative error is
                  $6.2\times10^{-16}$.}
                  \end{minipage}
                \label{fig:esfahan}
        \end{subfigure}
        \begin{subfigure}[b]{\textwidth}
                \centering
                \includegraphics[width=0.9\textwidth]{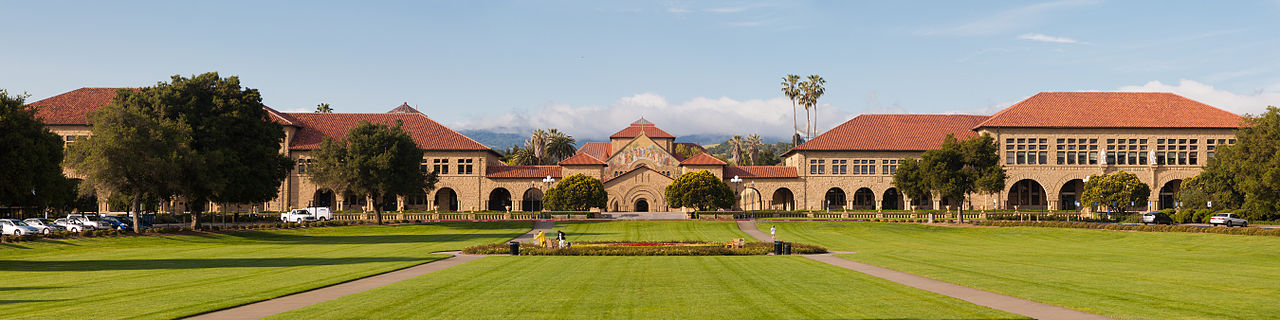}
                \begin{minipage}[t]{5in}\caption{Stanford main quad. Image size is $320\times
                  1280$ pixels; timing is 181.8120 sec or about
                  $20,700$ FFT units. The relative error is
                  $3.5 \times 10^{-14}$.}
                  \end{minipage}
                \label{fig:stanford}
        \end{subfigure}
        \begin{subfigure}[b]{\textwidth}
                \centering
                \includegraphics[width=0.9\textwidth]{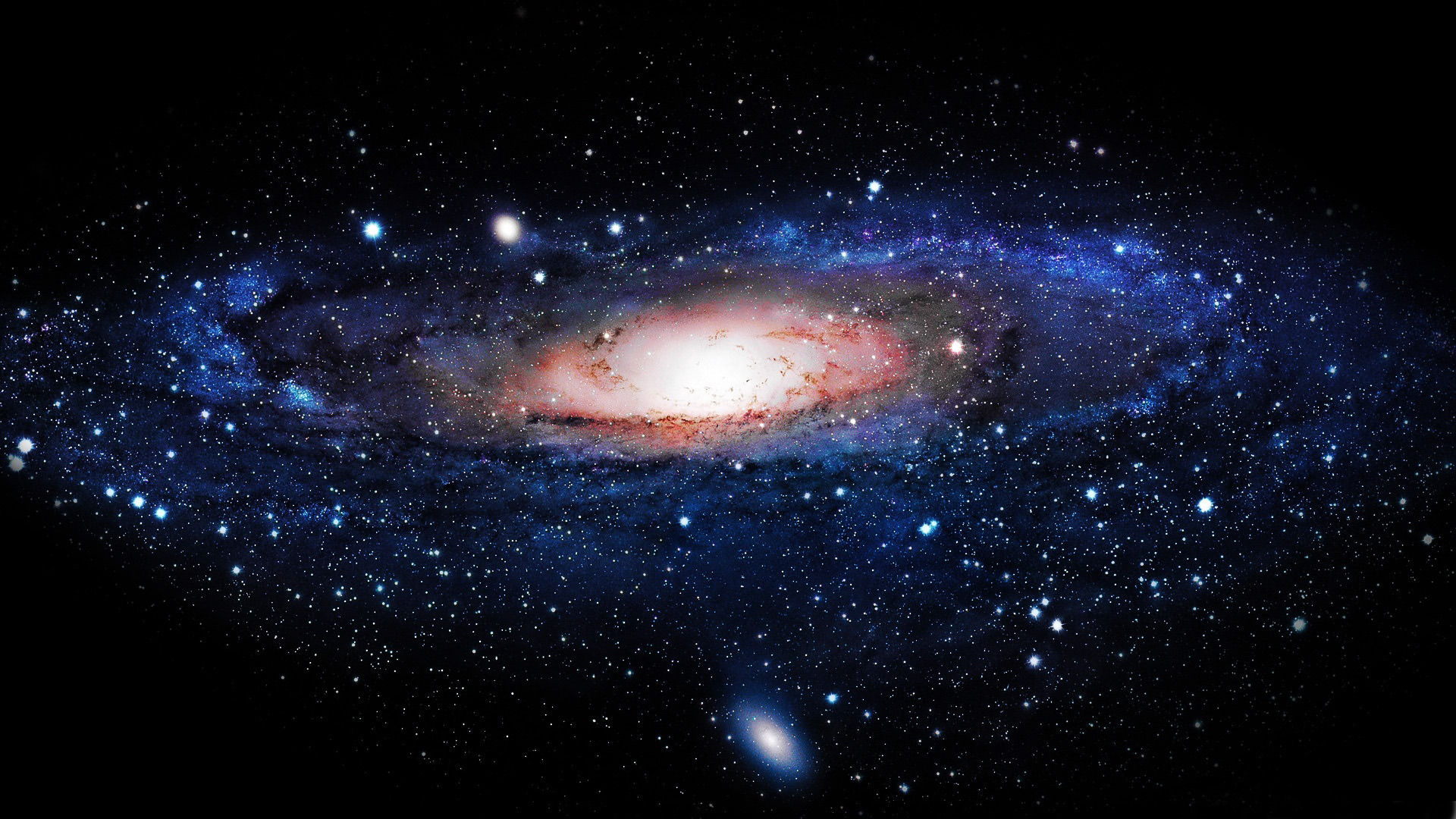}
                \begin{minipage}[t]{5in}
\caption{Milky way Galaxy. Image size is $1080\times
                  1920$ pixels; timing is 1318.1 sec or $41,900$ FFT
                  units. The relative error is $9.3 \times10^{-16}$.}
\end{minipage}
                \label{fig:milkyway}
              \end{subfigure}

              \caption{
                Performance of the WF algorithm on three scenic
                images. Image size, computational time in seconds and
                in units of FFTs are reported, as well as the relative
                error after 300 WF iterations.}
          \label{fig:MT}
\end{figure}

Figure \ref{fig:MT} shows the images recovered via the Wirtinger flow
algorithm. In all cases, WF gets 12 or 13 digits of precision in a
matter of minutes. To convey an idea of timing that is
platform-independent, we also report time in units of FFTs; one FFT
unit is the amount of time it takes to perform a single FFT on an
image of the same size.  Now all the workload is dominated by matrix
vector products of the form $\mtx{A} \vct{z}$ and $\mtx{A}^*
\vct{v}$. In details, each iteration of the power method in the
initialization step, or each update \eqref{graddescent} requires $40$
FFTs; the factor of 40 comes from the fact that we have 20 patterns
and that each iteration involves one FFT and one adjoint or inverse
FFT. Hence, the total number of FFTs is equal to
\[
20 \text{ patterns } \times 2 \text{ (one FFT and one IFFT) }
\times (300 \text{ gradient steps} + 50 \text{ power iterations}) =
14,000.
\]
Another way to state this is that the total workload of our algorithm
is roughly equal to $350$ applications of the sensing matrix $\mtx{A}$
and its adjoint $\mtx{A}^*$. For about 13 digits of accuracy (relative
error of about $10^{-13}$), Figure \ref{fig:MT} shows that we need
between 21,000 and 42,000 FFT units. This is within a factor between
1.5 and 3 of the optimal number computed above. This increase has to
do with the fact that in our implementation, certain variables are
copied into other temporary variables and these types of operations
cause some overhead. This overhead is non-linear and becomes more
prominent as the size of the signal increases.

For comparison, SDP based solutions such as
PhaseLift\cite{candes2012phaselift, candes2013phase} and
PhaseCut\cite{waldspurger2012phase} would be prohibitive on a laptop
computer as the lifted signal would not fit into memory. In the SDP
approach an $n$ pixel image become an $n^2/2$ array, which in the
first example already takes storing the lifted signal even for the
smallest image requires $(189\times 768)^2 \times 1/2 \times 8$ Bytes,
which is approximately 85 GB of space. (For the image of the Milky
Way, storage would be about 17 TB.) These large memory requirements
prevent the application of full-blown SDP solvers on desktop
computers.

\subsection{3D molecules}

\begin{figure}
        \centering
         \begin{subfigure}[b]{0.48\textwidth}
                \centering
                \includegraphics[scale=0.3]{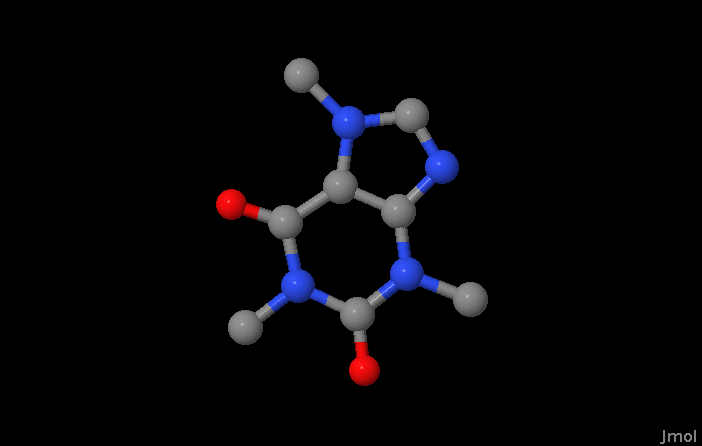}
                \caption{Schematic representation}
                \label{fig:cafschem}
        \end{subfigure}
        \begin{subfigure}[b]{0.48\textwidth}
                \centering
                \includegraphics[scale=0.45]{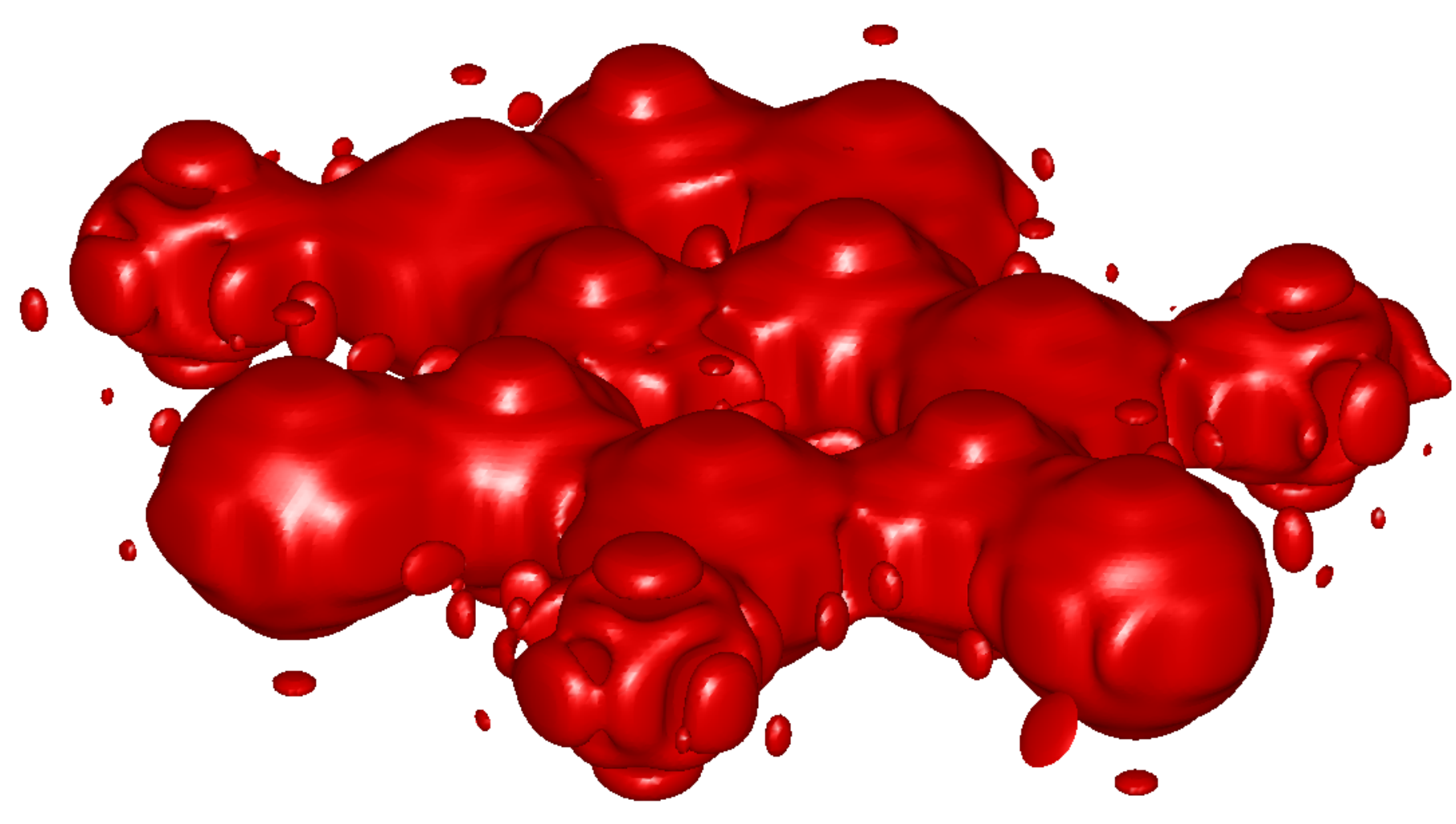}
                \caption{Electron density map}
                \label{fig:cafelec}
        \end{subfigure}

        \caption{Schematic representation and electron density map of the Caffeine molecule.}
          \label{fig:cafmol}
\end{figure}

\begin{figure}
        \centering
         \begin{subfigure}[b]{0.48\textwidth}
                \centering
                \includegraphics[scale=0.3]{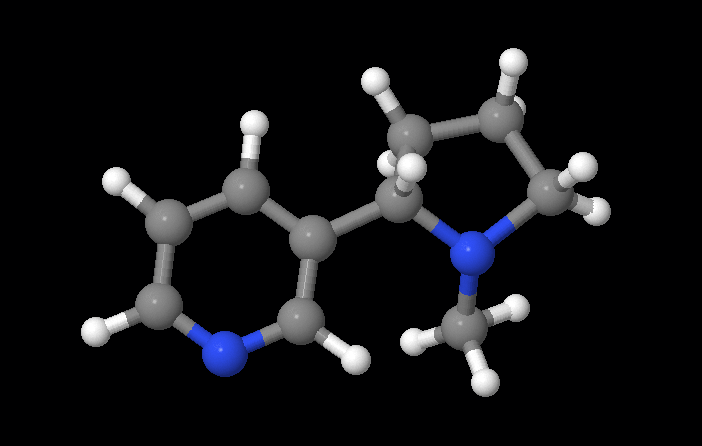}
                \caption{Schematic representation}
                \label{fig:nicschem}
        \end{subfigure}
        \begin{subfigure}[b]{0.48\textwidth}
                \centering
                \includegraphics[scale=0.45]{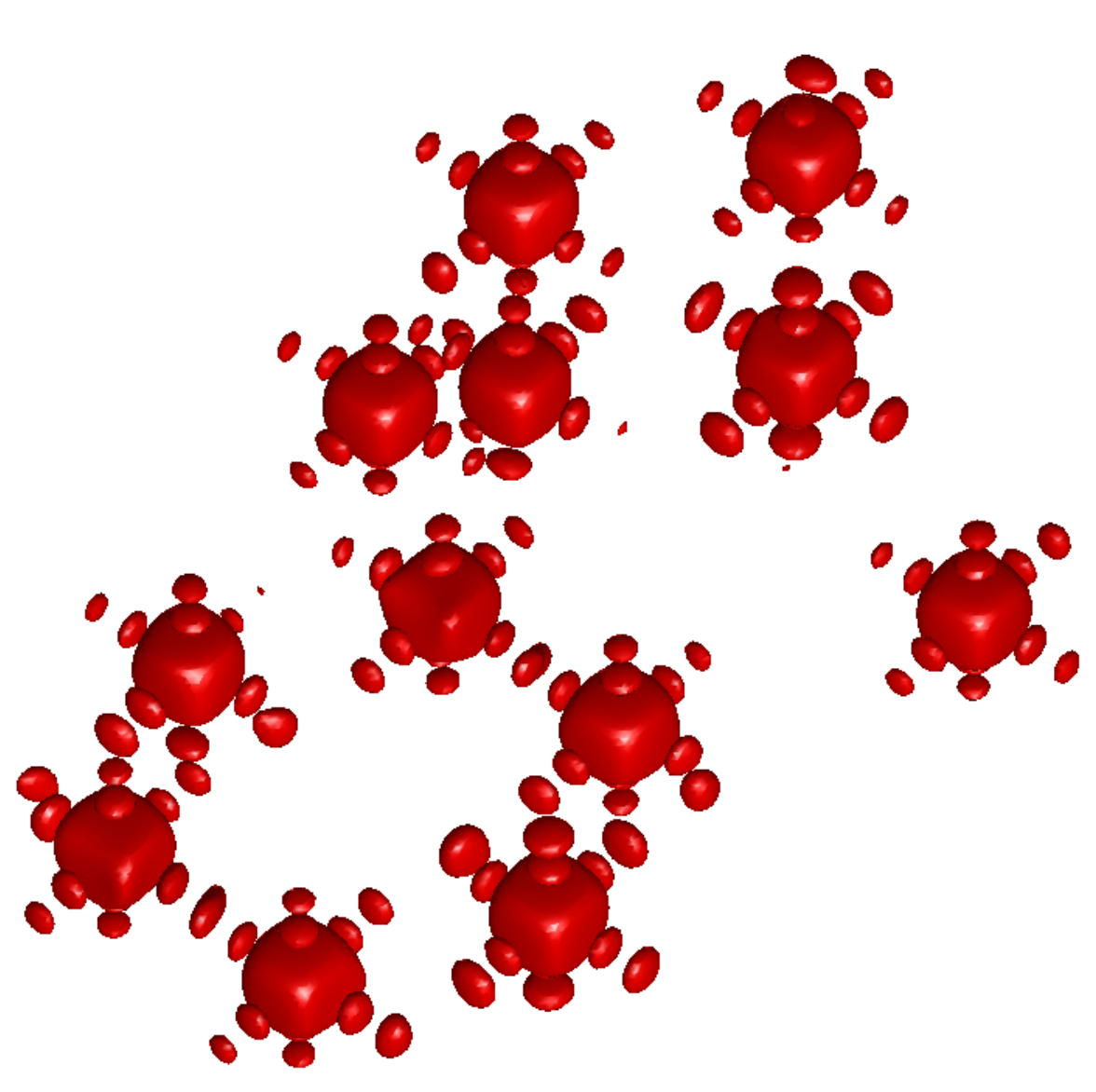}
                \caption{Electron density map}
                \label{fig:nicelec}
        \end{subfigure}
        \caption{Schematic representation and electron density map of the Nicotine molecule.}
          \label{fig:nicmol}
\end{figure}

Understanding molecular structure is a great contemporary scientific
challenge, and several techniques are currently employed to produce 3D
images of molecules; these include electron microscopy and X-ray
imaging. In X-ray imaging, for instance, the experimentalist
illuminates an object of interest, e.g.~a molecule, and then collects
the intensity of the diffracted rays, please see Figure \ref{fig:xray}
for an illustrative setup. Figures \ref{fig:cafmol} and
\ref{fig:nicmol} show the schematic representation and the
corresponding electron density maps for the Caffeine and Nicotine
molecules: the density map $\rho(x_1, x_2, x_3)$ is the 3D object we
seek to infer.  In this paper, we do not go as far 3D reconstruction
but demonstrate the performance of the Wirtinger flow algorithm for recovering
projections of 3D molecule density maps from simulated data.  For
related simulations using convex schemes we refer the reader to
\cite{fogel2013phase}.
\begin{figure} \centering
\begin{tikzpicture}
\node at (0,0) {\includegraphics[width=0.7\linewidth]{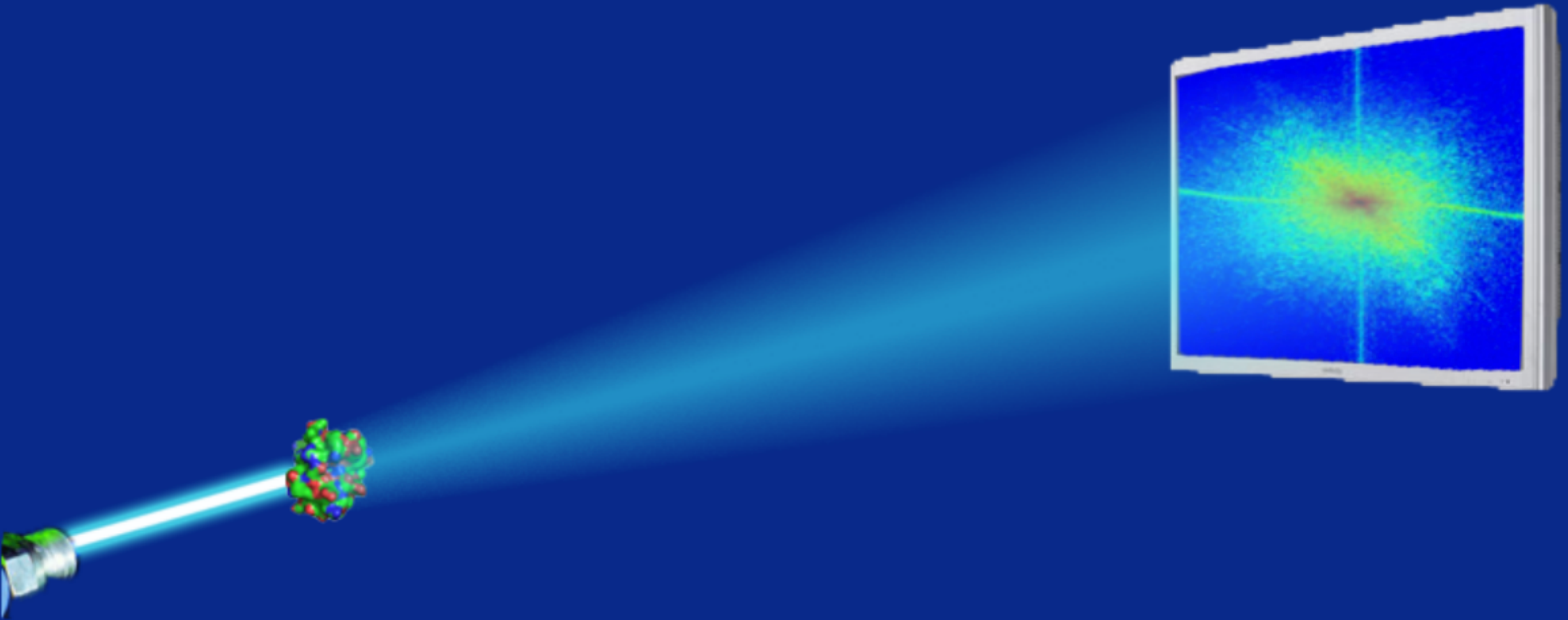}};
\node[white] at (-5.2,-1.3) {source};
\node[white] at (-3.3,-1.8) {molecule};
\node[white] at (4,-0.8) {diffraction patterns};
\end{tikzpicture}
\label{fig:xray}
\caption{An illustrative setup of diffraction patterns.}
\label{fig:xray}
\end{figure}

\begin{figure}
        \centering
                \includegraphics[scale=0.3]{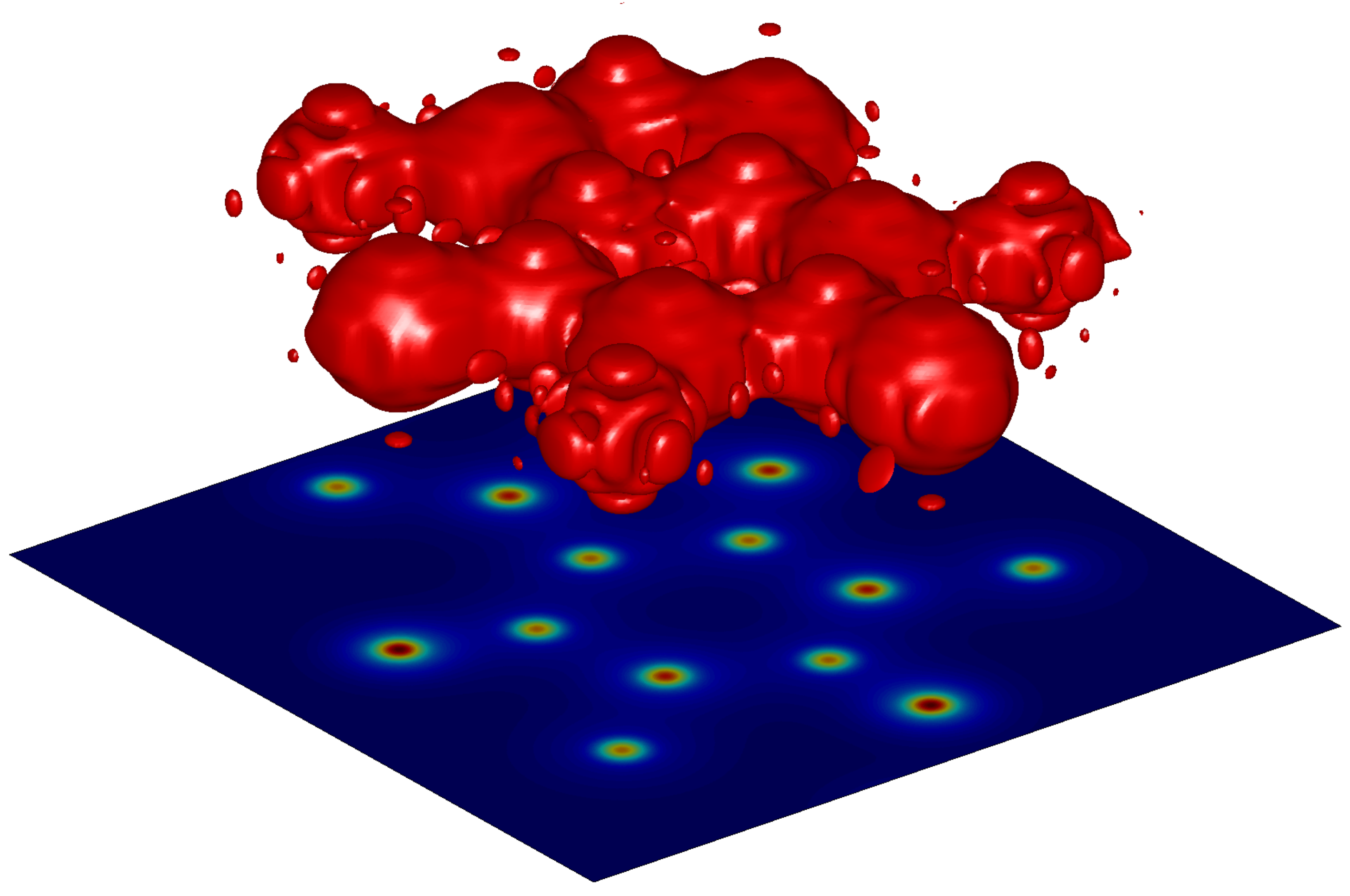}
                \caption{Electron density $\rho(x_1,x_2,x_3)$ of the
                  Caffeine molecule along with its projection onto the
                  $x_1x_2$-plane.}
          \label{fig:molwithproj}
\end{figure}

To derive signal equations, consider an experimental apparatus as in
Figure \ref{fig:xray}. If we imagine that light propagates in the
direction of the $x_3$-axis, an approximate model for the collected
data reads
\begin{align*}
  I(f_1,f_2) = \abs{ \int\left(\int \rho(x_1,x_2,x_3)
      \d x_3\right)e^{-2i\pi(f_1x_1+f_2x_2)}\d x_1 \d x_2}^2. 
\end{align*}
In other words, we collect the intensity of the diffraction pattern of
the projection $\int \rho(x_1,x_2,x_3) \d x_3$. The 2D image we wish
to recover is thus the line integral of the density map along a given
direction. As an example, the Caffeine molecule along with its
projection on the $x_1x_2$-plane (line integral in the $x_3$
direction) is shown in Figure \ref{fig:molwithproj}. Now, if we let
$R$ be the Fourier transform of the density $\rho$, one can re-express
the identity above as
\[
I(f_1,f_2)=\abs{R(f_1,f_2,0)}^2.
\]
Therefore, by imputing the missing phase using phase retrieval
algorithms, one can recover a slice of the 3D Fourier transfom of the
electron density map, i.e.~$R(f_1,f_2,0)$. Viewing the object from
different angles or directions gives us different slices. In a second
step we do not perform in this paper, one can presumably recover the
3D Fourier transform of the electron density map from all these slices
(this is the tomography or blind tomography problem depending upon
whether or not the projection angles are known) and, in turn, the 3D
electron density map.

\begin{figure}
        \centering
         \begin{subfigure}[b]{0.49\textwidth}
                \centering
                \captionsetup{justification=centering}
                \includegraphics[width=175pt]{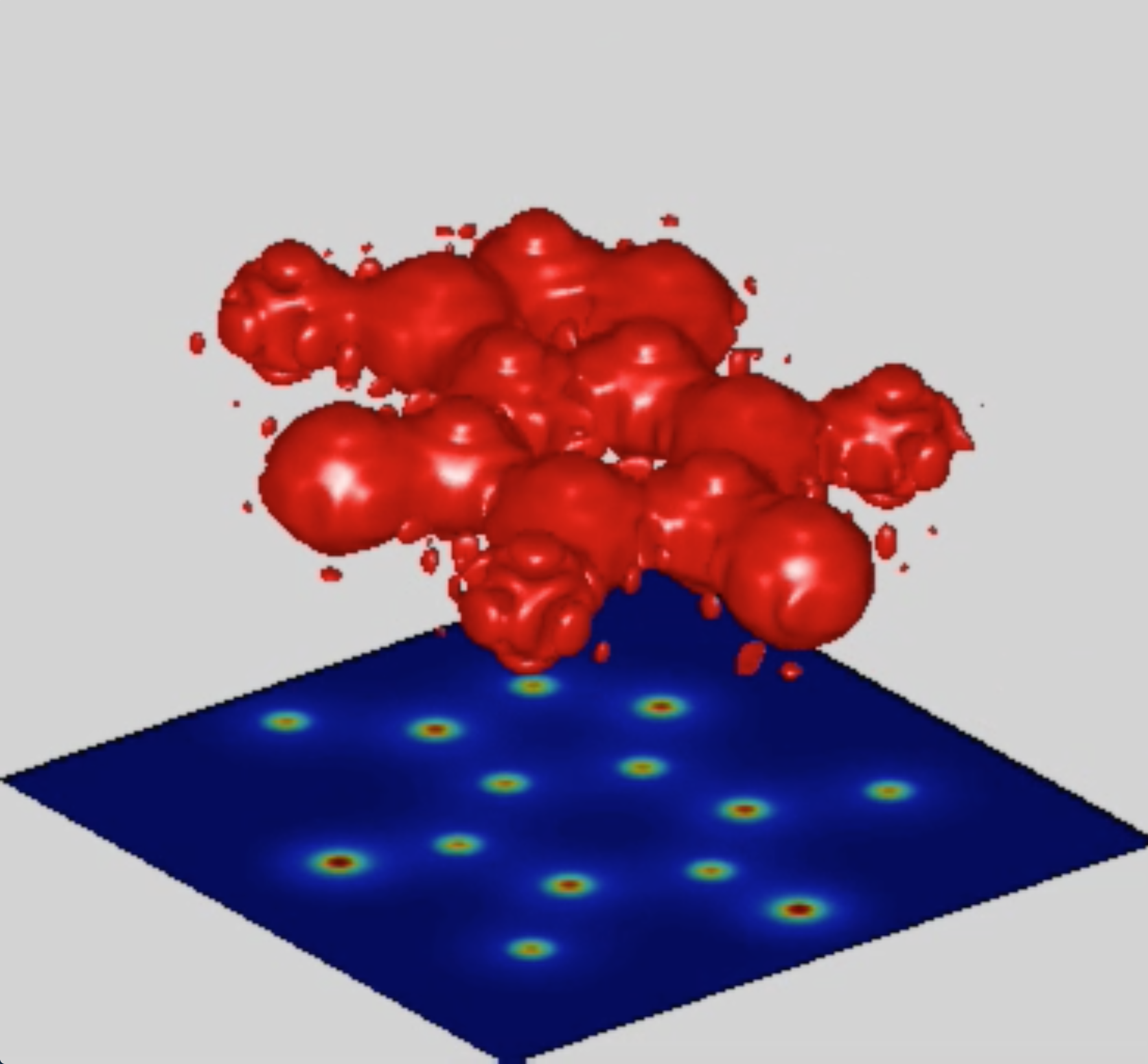}

                \caption{Caffeine molecule\\Mean rel.~error is
                  $9.6\times 10^{-6}$\\ Total time is $5.4$ hours }
               
                \label{fig:cafmov}
        \end{subfigure}
        \begin{subfigure}[b]{0.49\textwidth}
        \centering
        \captionsetup{justification=centering}
        \includegraphics[width=175pt]{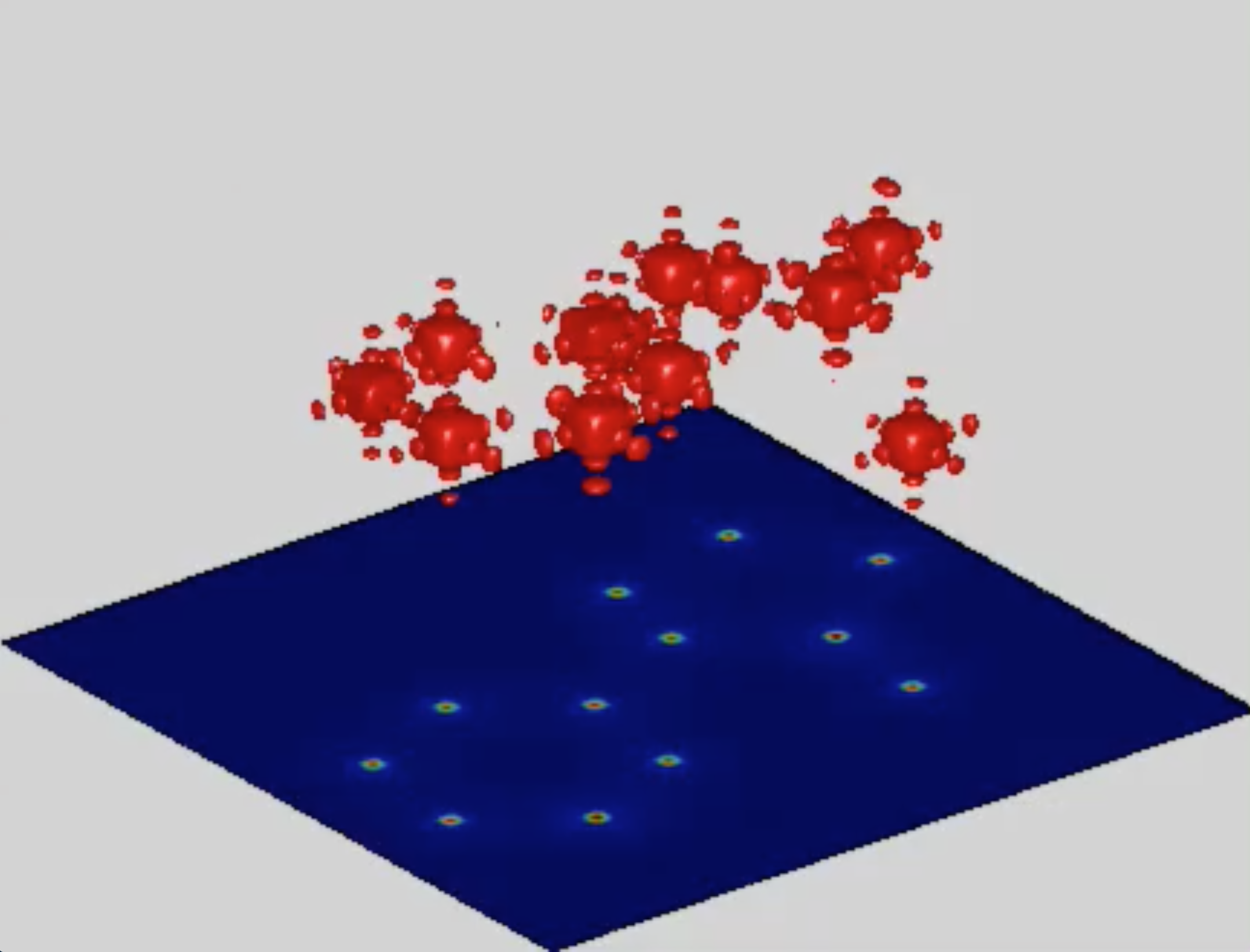}
\caption{Nicotine molecule\\Mean rel.~error is $1.7\times
  10^{-5}$\\Total time is $5.4$ hours }
                 \label{fig:nicmov}
        \end{subfigure}
        \caption{Reconstruction sequence of the projection of the
          Caffeine and Nicotine molecules along different
          directions. To see the videos please download and open the
          PDF file using Acrobat Reader.}
          \label{fig:mov}
\end{figure}

We now generate $51$ observation planes by rotating the $x_1x_2$-plane
around the $x_1$-axis by equally spaced angles in the interval
$[0,2\pi]$. Each of these planes is associated with a 2D projection of
size $1024\times 1024$, giving us 20 coded diffraction octanary
patterns (we use the same patterns for all 51 projections). We run the
Wirtinger flow algorithm with exactly the same parameters as in the previous
section, and stop after 150 gradient iterations.  Figure \ref{fig:mov}
reports the average relative error over the 51 projections and the
total computational time required for reconstructing all 51
images. 

\section{Theory for the Coded Diffraction Model}
\label{sec:Fourier}

We complement our study with theoretical results applying to the model
of coded diffraction patterns.  These results concern a variation of
the Wirtinger flow algorithm: whereas the iterations are exactly the same as
\eqref{graddescent}, the initialization applies an iterative scheme
which uses fresh sets of sample at each iteration. This is described
in Algorithm \ref{QIINITn}. In the CDP model, the partitioning assigns
to the same group all the observations and sampling vectors
corresponding to a given realization of the random code. This is
equivalent to partitioning the random patterns into $B+1$ groups. As a
result, sampling vectors in distinct groups are stochastically
independent.
\begin{algorithm}[h]
  \caption{Initialization via resampled Wirtinger Flow} 
\begin{algorithmic}
  \REQUIRE{Observations $\{y_r\} \in \R^m$ and number of blocks $B$.}
  \STATE \textbf{Partition} the observations and sampling vectors
  $\{y_r\}_{r=1}^m$ and $\{\vct{a}_r\}_{r=1}^m$ into $B+1$ groups of
  size $m' = \lfloor m/(B+1) \rfloor$. For each group $b = 0, 1,
  \ldots, B$, set
  \begin{align*}
    f(\vct{z};b)=\frac{1}{2m'}\sum_{r=1}^{m'}\bigg(y_r^{(b)}-\abs{\langle
      \vct{a}_r^{(b)},\vct{z}\rangle}^2\bigg)^2,
  \end{align*}
  where $\{\vct{a}_r^{(b)}\}$ are those sampling vectors belonging to
  the $b$th group (and likewise for $\{y_r^{(b)}\}$).
      \STATE \textbf{Initialize} $\tilde{\vct{u}}_0$ to be eigenvector
      corresponding to the largest eigenvalue of
  \begin{equation*}
  \mtx{Y}=\frac{1}{m'}\sum_{r=1}^{m'}y_r^{(0)}\vct{a}_r^{(0)}{\vct{a}_r^{(0)}}^*
  \end{equation*}
normalized as in Algorithm \eqref{QIINIT}. 
\STATE \textbf{Loop}: \FOR {$b= 0$ \TO $B-1$}
\STATE 
\[
 \vct{u}_{b+1}=\vct{u}_b-\frac{\tilde{\mu}}{\twonorm{\vct{u}_0}^2}\nabla f(\vct{u}_b;b)
\]
  \ENDFOR
  \ENSURE{$\vct{z}_0=\vct{u}_{B}$.}
\end{algorithmic}
\label{QIINITn}
\end{algorithm}

\begin{theorem}
\label{CDPthm} 
Let $\vct{x}$ be an arbitrary vector in $\C^n$ and assume we collect
$L$ admissible coded diffraction patterns with $L\ge c_0 \cdot
\left(\log n\right)^4$, where $c_0$ is a sufficiently large numerical
constant. Then the initial solution $\vct{z}_0$ of Algorithm \ref{QIINITn}\footnote{We choose the number of partitions $B$ in Algorithm \ref{QIINITn} to obey $B\ge c_1\log n$ for $c_1$ a sufficiently large numerical
constant.} obeys
\begin{align}
\label{CDPinit}
\text{\em dist}(\vct{z}_0,\vct{x}) \le \frac{1}{8\sqrt{n}}\twonorm{\vct{x}}
\end{align}
with probability at least $1-(4L+2)/{n^3}$. Further, take a constant
learning parameter sequence, $\mu_\tau= \mu$ for all $\tau = 1, 2,
\ldots$ and assume $\mu \le c_1$ for some fixed numerical constant
$c_1$.  Then there is an event of probability at least
$1-(2L+1)/{n^3}-{1}/{n^2}$, such that on this event, starting from any
initial solution $\vct{z}_0$ obeying \eqref{CDPinit}, we have
\begin{equation}
\label{concCDP}
  \text{\em dist}(\vct{z}_\tau,\vct{x})\le
  \frac{1}{8\sqrt{n}}\left(1-\frac{\mu}{3}\right)^{\tau/2} \cdot
  \twonorm{\vct{x}}.
\end{equation} 
In the Gaussian model, both statements also hold with high probability
provided that $m\ge c_0 \cdot n \, (\log n)^2$, where $c_0$ is a
sufficiently large numerical constant.
\end{theorem}

Hence, we achieve perfect recovery from on the order of $n (\log n)^4$
samples arising from a coded diffraction experiment. Our companion
paper \cite{PRCDP} established that PhaseLift---the SDP
relaxation---is also exact with a sampling complexity on the order of
$n (\log n)^4$ (this has recently been improved to $n (\log n)^2$
\cite{gross2014improved}). We believe that the sampling complexity of
both approaches (WF and SDP) can be further reduced to $n\log n$ (or
even $n$ for certain kind of patterns). We leave this to future
research.

Setting $\mu=c_1$ yields $\epsilon$ accuracy in $\mathcal{O}(\log
1/\epsilon)$ iterations. As the computational work at each iteration
is dominated by two matrix-vector products of the form $\mtx{A}
\vct{z}$ and $\mtx{A}^* \vct{v}$, it follows that the overall
computational is at most $\mathcal{O}(nL\log n\log{1}/{\epsilon})$. In
particular, this approach yields a near-linear time algorithm in the
CDP model (linear in the dimension of the signal $n$). In the Gaussian
model, the complexity scales like $\mathcal{O}(mn\log{1}/{\epsilon})$.

\section{Wirtinger Derivatives}
\label{wirt}

Our gradient step \eqref{graddescent} uses a notion of derivative,
which can be interpreted as a Wirtinger derivative. The purpose of
this section is thus to gather some results concerning Wirtinger
derivatives of real valued functions over complex variables. Here and
below, $\mtx{M}^T$ is the transpose of the matrix $\mtx{M}$, and
$\bar{c}$ denotes the complex conjugate of a scalar $c \in
\C$. Similarly, the matrix $\bar{\mtx{M}}$ is obtained by taking
complex conjugates of the elements of $\mtx{M}$.

Any complex-or real-valued function
\begin{align*}
f(\vct{z})=f(\vct{x},\vct{y})=u(\vct{x},\vct{y})+iv(\vct{x},\vct{y})
\end{align*}
of several complex variables can be written in the form
$f(\vct{z},\bar{\vct{z}})$, where $f$ is holomorphic in
$\vct{z}=\vct{x}+i\vct{y}$ for fixed $\bar{\vct{z}}$ and holomorphic
in $\bar{\vct{z}}=\vct{x}-i\vct{y}$ for fixed $\vct{z}$. This holds as
long as a the real-valued functions $u$ and $v$ are differentiable as
functions of the real variables $\vct{x}$ and $\vct{y}$.  As an
example, consider
\begin{align*}
  f(\vct{z})=\bigl(y-\abs{\vct{a}^*\vct{z}}^2\bigr)^2=(y-{\vct{\bar{z}}}^T\vct{a}\vct{a}^*\vct{z})^2=f(\vct{z},\vct{\bar{z}}).
\end{align*}
with $\vct{z}, \vct{a}\in\C^n$ and $y\in\R$. While $f(\vct{z})$ is not
holomorphic in $\vct{z}$, $f(\vct{z},\vct{\bar{z}})$ is holomorphic in
$\vct{z}$ for a fixed $\bar{\vct{z}}$, and vice versa.

This fact underlies the development of the \emph{Wirtinger
  calculus}. In essence, the {\em conjugate coordinates}
\begin{align*}
  \begin{bmatrix}\vct{z}\\\bar{\vct{z}}\end{bmatrix}\in\C^n\times\C^n,\quad\vct{z}=\vct{x}+i\vct{y}\quad\text{and}\quad\bar{\vct{z}}=\vct{x}-i\vct{y},
\end{align*}
can serve as a formal substitute for the representation $(\vct{x},
\vct{y}) \in\R^{2n}$.  This leads to the following derivatives
\begin{align*}
  \frac{\partial f}{\partial\vct{z}}:=\frac{\partial f(\vct{z},\bar{\vct{z}})}{\partial \vct{z}}|_{\bar{\vct{z}}=\text{constant}}= \left[\frac{\partial f}{\partial\vct{z}_1},\frac{\partial f}{\partial\vct{z}_2},\ldots,\frac{\partial f}{\partial\vct{z}_n}\right]_{\bar{\vct{z}}=\text{constant}},\\
  \frac{\partial f}{\partial\bar{\vct{z}}}:=\frac{\partial
    f(\vct{z},\bar{\vct{z}})}{\partial
    \bar{\vct{z}}}|_{\vct{z}=\text{constant}}=\left[\frac{\partial
      f}{\partial\bar{\vct{z}}_1},\frac{\partial
      f}{\partial\bar{\vct{z}}_2},\ldots,\frac{\partial
      f}{\partial\bar{\vct{z}}_n}\right]_{\vct{z}=\text{constant}}.
\end{align*}
Our definitions follow standard notation from multivariate calculus so
that derivatives are row vectors and gradients are column vectors. In
this new coordinate system the complex gradient is given by
\begin{align*}
  \nabla_c f = \left[\frac{\partial
      f}{\partial\vct{z}}, \frac{\partial
      f}{\partial\bar{\vct{z}}}\right]^*.
\end{align*}
Similarly, we define
\begin{align*}
\mathcal{H}_{\vct{z}\vct{z}}:=\frac{\partial}{\partial \vct{z}}\left(\frac{\partial f}{\partial \vct{z}}\right)^*,\quad\mathcal{H}_{\bar{\vct{z}}\vct{z}}:=\frac{\partial}{\partial \bar{\vct{z}}}\left(\frac{\partial f}{\partial \vct{z}}\right)^*,\quad\mathcal{H}_{\vct{z}\bar{\vct{z}}}:=\frac{\partial}{\partial \vct{z}}\left(\frac{\partial f}{\partial \bar{\vct{z}}}\right)^*,\quad\mathcal{H}_{\bar{\vct{z}}\bar{\vct{z}}}:=\frac{\partial}{\partial \bar{\vct{z}}}\left(\frac{\partial f}{\partial \bar{\vct{z}}}\right)^*.
\end{align*}
In this coordinate system the complex Hessian is given by
\begin{align*}
\nabla^2 f:=\begin{bmatrix}\mathcal{H}_{\vct{z}\vct{z}} & \mathcal{H}_{\bar{\vct{z}}\vct{z}} \\ \mathcal{H}_{\vct{z}\bar{\vct{z}}} & \mathcal{H}_{\bar{\vct{z}}\bar{\vct{z}}} \end{bmatrix}.
\end{align*}
Given vectors $\vct{z}$ and $\Delta\vct{z}\in\C^n$, we have defined
the gradient and Hessian in a manner such that Taylor's approximation
takes the form
\[
f(\vct{z}+\Delta\vct{z}) \approx f(\vct{z})\, + \, \left(\nabla_c
  f(\vct{z})\right)^*\begin{bmatrix}\Delta\vct{z}\\\text{ }\vspace{-0.1in}\\\overline{\Delta\vct{z}}\end{bmatrix}\,
+\, \frac{1}{2} \begin{bmatrix}\Delta\vct{z}\\\text{ }\vspace{-0.1in}\\\overline{\Delta\vct{z}}\end{bmatrix}^*
\nabla^2f
(\vct{z}) \begin{bmatrix}\Delta\vct{z}\\\text{ }\vspace{-0.1in}\\\overline{\Delta\vct{z}}\end{bmatrix}.
\]
If we were to run gradient descent in this new coordinate system, the
iterates would be
\begin{align}
\label{tempcgrad}
\begin{bmatrix}\vct{z}_{\tau+1}\\\bar{\vct{z}}_{\tau+1}\end{bmatrix}=\begin{bmatrix}\vct{z}_\tau\\\bar{\vct{z}}_\tau\end{bmatrix}-\mu\begin{bmatrix}
  \big({\partial
    f}/{\partial\vct{z}}\big)^*|_{\vct{z}=\vct{z}_\tau}\\
  \big({\partial
    f}/{\partial\bar{\vct{z}}}\big)^*|_{\vct{z}=\vct{z}_\tau}\end{bmatrix}
\end{align}
Note that when $f$ is a real-valued function (as in this paper) we have 
\begin{align*}
\overline{\frac{\partial f}{\partial\vct{z}}}=\frac{\partial f}{\partial\bar{\vct{z}}}.
\end{align*}
Therefore, the second set of updates in \eqref{tempcgrad} is just the
conjugate of the first. Thus, it is sufficient to keep track of the
first update, namely,
\begin{align*}
\vct{z}_{\tau+1}=\vct{z}_\tau-\mu\left({\partial f}/{\partial\vct{z}}\right)^*.
\end{align*}
For real valued functions of complex variables, setting 
\begin{align*}
\nabla f(\vct{z})=\left(\frac{\partial f}{\partial\vct{z}}\right)^*
\end{align*}
gives the gradient update 
\begin{align*}
\vct{z}_{\tau+1}=\vct{z}_\tau-\mu\nabla f(\vct{z}_\tau).
\end{align*}

The reader may wonder why we choose to work with conjugate coordinates
as there are alternatives: in particular, we could view the complex
variable $\vct{z}=\vct{x}+i\vct{y}\in\C^n$ as a vector in $\R^{2n}$
and just run gradient descent in the $\vct{x},\vct{y}$ coordinate
system. The main reason why conjugate coordinates are particularly
attractive is that expressions for derivatives become significantly
simpler and resemble those we obtain in the real case, where $f: \R^n
\rightarrow \R$ is a function of real variables.

\section{Proofs}
\label{proofsec}

\subsection{Preliminaries}

We first note that in the CDP model with admissible CDPs
$\twonorm{\vct{a}_r}\le\sqrt{6n}$ for all $r=1,2,\ldots,m$, as in our CDP model $\abs{d}\le\sqrt{3}<\sqrt{6}$. In the
Gaussian model the measurements vectors also obey
$\twonorm{\vct{a}_r}\le\sqrt{6n}$ for all $r=1,2,\ldots,m$ with
probability at least $1-me^{-1.5n}$. Throughout the proofs, we assume
we are on this event without explicitly mentioning it each time.

Before we begin with the proofs we should mention that we will prove
our result using the update
\begin{align}
\label{idealizedupdate}
\vct{z}_{\tau+1}=\vct{z}_\tau-\frac{\mu}{\twonorm{\vct{x}}^2}\nabla
f(\vct{z}_\tau),
\end{align}
in lieu of the WF update
\begin{align}
\label{QIupdate}
\vct{z}_{\tau+1}=\vct{z}_\tau-\frac{\mu_{\text{WF}}}{\twonorm{\vct{z}_0}^2}\nabla f(\vct{z}_\tau).
\end{align}
Since $\abs{\twonorm{\vct{z}_0}^2-\twonorm{\vct{x}}^2} \le
\frac{1}{64}\twonorm{\vct{x}}^2$ holds with high probability as proven
in Section \ref{initproofsec}, we have
\begin{align}
\label{ineqmu}
{\twonorm{\vct{z}_0}^2}\ge\frac{63}{64}\, {\twonorm{\vct{x}}^2}.
\end{align}
Therefore, the results for the update \eqref{idealizedupdate}
automatically carry over to the WF update with a simple rescaling of
the upper bound on the learning parameter. More precisely, if we prove
that the update \eqref{idealizedupdate} converges to a global optimum
as long as $\mu\le\mu_0$, then the convergence of the WF update to a
global optimum is guaranteed as long as $\mu_{\text{WF}}\le
\frac{63}{64}\mu_0$. Also, the update in \eqref{idealizedupdate} is
invariant to the Euclidean norm of $\vct{x}$. Therefore, without loss
of generality we will assume throughout the proofs that
$\twonorm{\vct{x}}=1$.

We remind the reader that throughout $\vct{x}$ is a solution to our
quadratic equations, i.e.~obeys $y = |\mtx{A} \vct{x}|^2$ and that the
sampling vectors are independent from $\vct{x}$. Define 
\[
P:=\{\vct{x}e^{i \phi}: \phi\in[0,2\pi]\}.
\]
to be the set of all vectors that differ from the planted solution
$\vct{x}$ only by a global phase factor.  We also introduce the set of
all points that are close to $P$, 
\begin{align}
\label{Eepsilon}
E(\epsilon):=\{\vct{z} \in \mathbb{C}^n:\text{dist}(\vct{z},P) \leq \epsilon \},
\end{align}
Finally for any vector $\vct{z}\in\C^n$ we define the phase
$\phi(\vct{z})$ as
\begin{align*}
\phi(\vct{z}):=\underset{\phi\in[0,2\pi]}{\arg\min}\quad\twonorm{\vct{z}-e^{i\phi}\vct{x}},
\end{align*}
so that
\begin{align*}
\text{dist}(\vct{z},\vct{x})=\twonorm{\vct{z}-e^{i\phi(\vct{z})}\vct{x}}.
\end{align*}

\subsection{Formulas for the complex gradient and Hessian}
\label{gradHess}

We gather some useful gradient and Hessian calculations that will be
used repeatedly.  Starting with 
\begin{align*}
f(\vct{z})=\frac{1}{2m}\sum_{r=1}^m\left(y_r-\bar{\vct{z}}^T(\vct{a}_r\vct{a}_r^*)\vct{z}\right)^2=\frac{1}{2m}\sum_{r=1}^m\left(y_r-\vct{z}^T(\vct{a}_r\vct{a}_r^*)^T\bar{\vct{z}}\right)^2,
\end{align*}
we establish
\begin{align*}
\left(\frac{\partial}{\partial \vct{z}}f(\vct{z})\right)^T&=\frac{1}{m}\sum_{r=1}^m\left(\vct{z}^T(\vct{a}_r\vct{a}_r^*)^T\bar{\vct{z}} - y_r\right)(\vct{a}_r\vct{a}_r^*)^T\bar{\vct{z}}.
\end{align*}
This gives 
\begin{align*}
\nabla f(\vct{z})=\left(\frac{\partial}{\partial \vct{z}}f(\vct{z})\right)^* &=\frac{1}{m}\sum_{r=1}^m\left(\bar{\vct{z}}^T(\vct{a}_r\vct{a}_r^*)\vct{z}-y_r\right)(\vct{a}_r\vct{a}_r^*)\vct{z}.
\end{align*}

For the second derivative, we write
\[
\mathcal{H}_{\vct{z}\vct{z}}=\frac{\partial} {\partial\vct{z}}\left(\frac{\partial}{\partial \vct{z}}f(\vct{z})\right)^* =\frac{1}{m}\sum_{r=1}^m \left(2|\vct{a}_r^*\vct{z}|^2-y_r\right)\vct{a}_r\vct{a}_r^*
\]
and
\[
\mathcal{H}_{\bar{\vct{z}}\vct{z}}=\frac{\partial} {\partial\bar{\vct{z}}}\left(\frac{\partial}{\partial \vct{z}}f(\vct{z})\right)^*=\frac{1}{m}\sum_{r=1}^m (\vct{a}_r^*\vct{z})^2\vct{a}_r\vct{a}_r^T.
\]
Therefore,
\begin{align*}
\nabla^2 f(\vct{z})=\frac{1}{m}\sum_{r=1}^m\begin{bmatrix} (2\abs{\vct{a}_r^*\vct{z}}^2-y_r)\vct{a}_r\vct{a}_r^* & {(\vct{a}_r^*\vct{z})}^2\vct{a}_r\vct{a}_r^T \\ (\overline{\vct{a}_r^*\vct{z}})^2\bar{\vct{a}}_r\vct{a}_r^* & (2\abs{\vct{a}_r^*\vct{z}}^2-y_r)\bar{\vct{a}}_r\vct{a}_r^T  \end{bmatrix}.
\end{align*}

\subsection{Expectation and concentration}
\label{expandconsec}

This section gathers some useful intermediate results whose proofs are
deferred to Appendix \ref{proofexpconc}. The first two lemmas
establish the expectation of the Hessian, gradient and a related
random variable in both the Gaussian and admissible CDP
models.\footnote{In the CDP model the expectation is with respect to
  the random modulation pattern.}
\begin{lemma}
\label{expectHessian}
Recall that $\vct{x}$ is a solution obeying $\twonorm{\vct{x}}=1$,
which is independent from the sampling vectors. Furthermore, assume
the sampling vectors $\vct{a}_r$ are distributed according to either the
Gaussian or admissible CDP model. Then 
\[
\E[\nabla^2 f(\vct{x})]=\mtx{I}_{2n} +\frac{3}{2}\begin{bmatrix} \vct{x} \\ \bar{\vct{x}} \end{bmatrix}[\vct{x}^*,\vct{x}^T]-\frac{1}{2}\begin{bmatrix} \vct{x} \\-\bar{\vct{x}} \end{bmatrix} [\vct{x}^*, -\vct{x}^T].
\] 
\end{lemma}

\begin{lemma}
\label{expGrad}
In the setup of Lemma \ref{expectHessian}, let $\vct{z}\in\C^n$ be a
fixed vector independent of the sampling vectors. We have
\[
\E[\nabla f(\vct{z})]=(\mtx{I}-\vct{x}\vct{x}^*)\vct{z}+2\left(\twonorm{\vct{z}}^2-1\right)\vct{z}.
\] 
\end{lemma}
The next lemma gathers some useful identities in the Gaussian model.
\begin{lemma}
\label{usefulexpG}
Assume $\vct{u}, \vct{v}\in\C^n$ are fixed vectors obeying $\twonorm{\vct{u}}=\twonorm{\vct{v}}=1$ which are independent of the sampling vectors. Furthermore, assume the measurement vectors $\vct{a}_r$ are distributed according to the Gaussian model. Then 
\begin{align}
\label{term1}\E \big[\left(\Real(\vct{u}^*\vct{a}_r\vct{a}_r^*\vct{v})\right)^2\big]=&\frac{1}{2}+\frac{3}{2}(\Real(\vct{u}^*\vct{v}))^2-\frac{1}{2}(\Imag(\vct{u}^*\vct{v}))^2\\
\label{term2}\E \big[\Real(\vct{u}^*\vct{a}_r\vct{a}_r^*\vct{v})\abs{\vct{a}_r^*\vct{v}}^2\big]=&2\Real(\vct{u}^*\vct{v})\\
\label{term3}\E \big[\abs{\vct{a}_r^*\vct{v}}^{2k}\big]=&k!.
\end{align}
\end{lemma}

The next lemma establishes the concentration of the Hessian around its
mean for both the Gaussian and the CDP model.
\begin{lemma}
\label{concenHessian}
In the setup of Lemma \ref{expectHessian}, assume the vectors
$\vct{a}_r$ are distributed according to either the Gaussian or
admissible CDP model with a sufficiently large number of
measurements. This means that the number of samples obeys $m\ge
c(\delta)\cdot n\log n$ in the Gaussian model and the number of
patterns obeys $L \ge c(\delta)\cdot \log^3 n$ in the CDP model. Then
\begin{equation}
\left\|\nabla^2 f(\vct{x})-\mathbb{E}[\nabla^2 f(\vct{x})]\right\|\leq \delta,
\end{equation}
holds with probability at least $1-10e^{-\gamma n}-{8}/{n^2}$ and
$1-(2L+1)/{n^3}$ for the Gaussian and CDP models, respectively.
\end{lemma}

We will also make use of the two results below, which are corollaries
of the three lemmas above. These corollaries are also proven in
Appendix \ref{proofexpconc}.
\begin{corollary}
\label{corineq}
Suppose $\left\|\nabla^2 f(\vct{x})-\mathbb{E}[\nabla^2
  f(\vct{x})]\right\|\leq \delta$. Then for all $\vct{h}\in\C^n$
obeying $\twonorm{\vct{h}}=1$, we have
\begin{align*}
\frac{1}{m}\sum_{r=1}^m \operatorname{Re}\big(\vct{h}^*\vct{a}_r\vct{a}_r^*\vct{x}\big)^2&=\frac{1}{4}\sum_{r=1}^m\begin{bmatrix}\vct{h}\\\bar{\vct{h}}\end{bmatrix}^*\nabla^2 f(\vct{x})\begin{bmatrix}\vct{h}\\\bar{\vct{h}}\end{bmatrix}
\le \left(\frac{1}{2}\twonorm{\vct{h}}^2+\frac{3}{2}\operatorname{Re}(\vct{x}^*\vct{h})^2-\frac{1}{2}\operatorname{Im}(\vct{x}^*\vct{h})^2\right)+\frac{\delta}{2}.
\end{align*}
In the other direction,
\begin{align*}
\frac{1}{m}\sum_{r=1}^m \operatorname{Re}\big(\vct{h}^*\vct{a}_r\vct{a}_r^*\vct{x}\big)^2\ge \left(\frac{1}{2}\twonorm{\vct{h}}^2+\frac{3}{2}\operatorname{Re}(\vct{x}^*\vct{h})^2-\frac{1}{2}\operatorname{Im}(\vct{x}^*\vct{h})^2\right)-\frac{\delta}{2}.
\end{align*}
\end{corollary}

\begin{corollary}
\label{corfirst}
Suppose $\left\|\nabla^2 f(\vct{x})-\mathbb{E}[\nabla^2
  f(\vct{x})]\right\|\leq \delta$. Then for all $\vct{h}\in\C^n$
obeying $\twonorm{\vct{h}}=1$, we have
\begin{align*}
\frac{1}{m}\sum_{r=1}^m\abs{\vct{a}_r^*\vct{x}}^2\abs{\vct{a}_r^*\vct{h}}^2=\vct{h}^*\left(\frac{1}{m}\sum_{r=1}^m\abs{\vct{a}_r^*\vct{x}}^2\vct{a}_r\vct{a}_r^*\right)\vct{h}\ge(1-\delta)\twonorm{\vct{h}}^2+\abs{\vct{h}^*\vct{x}}^2\ge(1-\delta)\twonorm{\vct{h}}^2,
\end{align*}
and
\begin{align*}
\frac{1}{m}\sum_{r=1}^m\abs{\vct{a}_r^*\vct{x}}^2\abs{\vct{a}_r^*\vct{h}}^2=\vct{h}^*\left(\frac{1}{m}\sum_{r=1}^m\abs{\vct{a}_r^*\vct{x}}^2\vct{a}_r\vct{a}_r^*\right)\vct{h}\le(1+\delta)\twonorm{\vct{h}}^2+\abs{\vct{h}^*\vct{x}}^2\le(2+\delta)\twonorm{\vct{h}}^2.
\end{align*}
\end{corollary}

The next lemma establishes the concentration of the gradient around its
mean for both Gaussian and admissible CDP models.
\begin{lemma}
\label{concenGrad}
In the setup of Lemma \ref{concenHessian}, let $\vct{z}\in \C^n$ be a
fixed vector independent of the sampling vectors obeying
$\text{dist}(\vct{z},\vct{x})\le\frac{1}{2}$. Then
\[
\twonorm{\nabla f(\vct{z})-\E[\nabla f(\vct{z})]}\le \delta\cdot
\emph{dist}(\vct{z},\vct{x}).
\] 
holds with probability at least $1-20e^{-\gamma m}-{4m}/{n^4}$ in
the Gaussian model and $1-(4L+2)/{n^3}$ in the CDP
model.
\end{lemma}

We finish with a result concerning the concentration of the sample
covariance matrix.
\begin{lemma}
\label{concenCovariance}
In the setup of Lemma \ref{concenHessian}, 
\begin{align*}
\left\|\mtx{I}_n - {m}^{-1}\sum_{r=1}^m \vct{a}_r\vct{a}_r^*\right\|\leq \delta,
\end{align*}
holds with probability at least $1-2e^{-\gamma m}$ for the Gaussian
model and $1-{1}/{n^2}$ in the CDP model.  On this event,
\begin{align}
\label{corspec}
(1-\delta)\twonorm{\vct{h}}^2\le\frac{1}{m}\sum_{r=1}^m\abs{\vct{a}_r^*\vct{h}}^2\le(1+\delta)\twonorm{\vct{h}}^2 \quad \text{for all } \vct{h} \in \C^n.
\end{align}
\end{lemma}

\subsection{General convergence analysis}
\label{GCAsec}
We will assume that the function $f$ satisfies a regularity condition
on $E(\epsilon)$, which essentially states that the gradient of the
function is well behaved. We remind the reader that $E(\epsilon)$, as defined in \eqref{Eepsilon}, is the set of points that are close to the path of global minimizers.

\begin{condition}[Regularity Condition] We say that the function $f$ satisfies the regularity condition or $RC(\alpha,\beta,\epsilon)$ if for all vectors $\vct{z}\in E(\epsilon)$ we have
\begin{align}
\label{regcond}
\operatorname{Re}\left(\big\langle \nabla f(\vct{z}),\vct{z}-\vct{x}e^{i \phi(\vct{z})}\big\rangle\right)  \ge \frac{1}{\alpha}\emph{dist}^2(\vct{z},\vct{x})+\frac{1}{\beta}\twonorm{\nabla f(\vct{z})}^2.
\end{align}
\end{condition}

In the lemma below we show that as long as the regularity condition
holds on $E(\epsilon)$ then Wirtinger Flow starting from an initial
solution in $E(\epsilon)$ converges to a global optimizer at a
geometric rate. Subsequent sections shall establish that this property
holds.

\begin{lemma}
\label{convergence}
Assume that $f$ obeys RC$(\alpha,\beta,\epsilon)$ for all $\vct{z} \in
E(\epsilon)$. Furthermore, suppose $\vct{z}_0 \in E$, and assume
$0<\mu\le {2}/{\beta}$. Consider the following update
\[
  \vct{z}_{\tau+1}=\vct{z}_\tau-\mu\nabla f(\vct{z}_\tau).
\]
Then for all $\tau$ we have $\vct{z}_\tau\in E(\epsilon)$ and
\begin{align*}
\emph{dist}^2(\vct{z}_\tau,\vct{x})\le \left(1-\frac{2\mu}{\alpha}\right)^\tau \emph{dist}^2(\vct{z}_0,\vct{x}).
  \end{align*}
\end{lemma}

We note that for $\alpha\beta<4$, \eqref{regcond} holds with the direction of the inequality reversed.\footnote{One can see this by applying Cauchy-Schwarz and calculating the determinant of the resulting quadratic form.} Thus, if $RC(\alpha,\beta,\epsilon)$ holds, $\alpha$ and $\beta$ must obey $\alpha\beta\ge 4$. As a result, under the stated assumptions of Lemma \ref{convergence} above, the factor $1-2\mu/\alpha\ge1-4/(\alpha\beta)$ is non-negative.

\begin{proof}
  The proof follows a structure similar to related results in the
  convex optimization literature e.g.~\cite[Theorem
  2.1.15]{nesterov2004introductory}. However, unlike these classical
  results where the goal is often to prove convergence to a unique
  global optimum, the objective function $f$ does not have a unique
  global optimum. Indeed, in our problem, if $\vct{x}$ is solution,
  then $e^{i\phi} \vct{x}$ is also solution. Hence, proper
  modification is required to prove convergence results. 

  We prove that if $\vct{z}\in E(\epsilon)$ then for all $0<\mu\le
  {2}/{\beta}$
\[
  \vct{z}_+=\vct{z}-\mu\nabla f(\vct{z})
\]
obeys
\begin{align}
\label{iter}
\text{dist}^2(\vct{z}_+,\vct{x})\le\left(1-\frac{2\mu}{\alpha}\right)\text{dist}^2(\vct{z},\vct{x}).
\end{align}
Therefore, if $\vct{z}\in E(\epsilon)$ then we also have $\vct{z}_+\in
E(\epsilon)$. The lemma follows by inductively applying
\eqref{iter}. Now simple algebraic manipulations together with the
regularity condition \eqref{regcond} give 
\begin{align*}
\twonorm{\vct{z}_{+}-\vct{x}e^{i \phi(\vct{z})}}^2=&\twonorm{\vct{z}-\vct{x}e^{i \phi(\vct{z})}-\mu\nabla f(\vct{z})}^2\nonumber\\
=&\twonorm{\vct{z}-\vct{x}e^{i \phi(\vct{z})}}^2-2\mu \operatorname{Re}\left(\big\langle \nabla f(\vct{z}),\left(\vct{z}-\vct{x}e^{i \phi(\vct{z})}\right)\big\rangle\right)+\mu^2\twonorm{\nabla f(\vct{z})}^2
\\
\le& \twonorm{\vct{z}-\vct{x}e^{i \phi(\vct{z})}}^2 -2\mu\left(\frac{1}{\alpha}\twonorm{\vct{z}-\vct{x}e^{i \phi(\vct{z})}}^2+\frac{1}{\beta}\twonorm{\nabla f(\vct{z})}^2\right)+\mu^2\twonorm{\nabla f(\vct{z})}^2\\
=&\left(1-\frac{2\mu}{\alpha}\right)\twonorm{\vct{z}-\vct{x}e^{i \phi(\vct{z})}}^2+\mu\left(\mu-\frac{2}{\beta}\right)\twonorm{\nabla f(\vct{z})}^2\\
\le&\left(1-\frac{2\mu}{\alpha}\right)\twonorm{\vct{z}-\vct{x}e^{i \phi(\vct{z})}}^2,
\end{align*}
where the last line follows from $\mu\le {2}/{\beta}$. The definition
of $\phi(\vct{z}_+)$ gives
\begin{align*}
\twonorm{\vct{z}_{+}-\vct{x}e^{i \phi(\vct{z}_+)}}^2\le \twonorm{\vct{z}_{+}-\vct{x}e^{i \phi(\vct{z})}}^2,
\end{align*}
which concludes the proof.
\end{proof}

\subsection{Proof of the regularity condition}

For any $\vct{z}\in E(\epsilon)$, we need to show that
\begin{align}
\label{tempRG}
\operatorname{Re}\left(\big\langle \nabla f(\vct{z}),\vct{z}-\vct{x}e^{i \phi(\vct{z})}\big\rangle\right)  \ge \frac{1}{\alpha}\text{dist}^2(\vct{z},\vct{x})+\frac{1}{\beta}\twonorm{\nabla f(\vct{z})}^2.
\end{align} 
We prove this with $\delta=0.01$ by establishing that our gradient
satisfies the local smoothness and local curvature conditions defined
below. Combining both these two properties gives \eqref{tempRG}.
\begin{condition}[Local Curvature Condition] We say that the function
  $f$ satisfies the local curvature condition or
  $LCC(\alpha,\epsilon,\delta)$ if for all vectors $\vct{z}\in
  E(\epsilon)$,
\begin{align}
\label{LCC}
\operatorname{Re}\left(\big\langle \nabla f(\vct{z}),\vct{z}-\vct{x}e^{i \phi(\vct{z})}\big\rangle\right)  \ge \left(\frac{1}{\alpha}+\frac{(1-\delta)}{4}\right)\emph{dist}^2(\vct{z},\vct{x})+\frac{1}{10m}\sum_{r=1}^m\abs{\vct{a}_r^*(\vct{z}-e^{i\phi(\vct{z})}\vct{x})}^4.
\end{align}

\end{condition}
This condition essentially states that the function curves
sufficiently upwards (along most directions) near the curve of global
optimizers.
\begin{condition}[Local Smoothness Condition] We say that the function $f$ satisfies the local smoothness condition or $LSC(\beta,\epsilon,\delta)$ if for all vectors $\vct{z}\in E(\epsilon)$ we have
\begin{align}
\label{LSC}
\twonorm{\nabla f(\vct{z})}^2\le\beta\left(\frac{(1-\delta)}{4}\emph{dist}^2(\vct{z},\vct{x})+\frac{1}{10m}\sum_{r=1}^m\abs{\vct{a}_r^*(\vct{z}-e^{i\phi(\vct{z})}\vct{x})}^4\right).
\end{align}
\end{condition}
This condition essentially states that the gradient of the function is
well behaved (the function does not vary too much) near the curve of
global optimizers.

\subsection{Proof of the local curvature condition}

For any $\vct{z}\in E(\epsilon)$, we want to prove the local curvature
condition \eqref{LCC}. Recall that
\begin{align*}
\nabla f(\vct{z})=\frac{1}{m}\sum_{r=1}^m\left(\abs{\langle\vct{a}_r,\vct{z}\rangle}^2-y_r\right)(\vct{a}_r\vct{a}_r^*)\vct{z},
\end{align*} 
and define $\vct{h}:=e^{-i\phi(\vct{z})}\vct{z}-\vct{x}$. To establish \eqref{LCC} it suffices to prove that 
\begin{multline}
\label{tempha}
\frac{1}{m}\sum_{r=1}^m \left(2\operatorname{Re}(\vct{h}^*\vct{a}_r\vct{a}_r^*\vct{x})^2+3\operatorname{Re}(\vct{h}^*\vct{a}_r\vct{a}_r^*\vct{x})|\vct{a}_r^*\vct{h}|^2+|\vct{a}_r^*\vct{h}|^4\right)-\left(\frac{1}{10m}\sum_{r=1}^m\abs{\vct{a}_r^*\vct{h}}^4\right)\\
\geq \left(\frac{1}{\alpha}+\frac{(1-\delta)}{4}\right)
\twonorm{\vct{h}}^2,
\end{multline}
holds for all $\vct{h}$ satisfying $\Imag(\vct{h}^*\vct{x})=0$,
$\|\vct{h}\|_2\leq \epsilon$. Equivalently, we only need to prove that
for all $\vct{h}$ satisfying $\Imag(\vct{h}^*\vct{x})=0$,
$\|\vct{h}\|_2=1$ and for all $s$ with $ 0\leq s \leq \epsilon$,
\begin{align}
\label{tempha2}
\frac{1}{m}\sum_{r=1}^m \left(2\operatorname{Re}(\vct{h}^*\vct{a}_r\vct{a}_r^*\vct{x})^2+3s\operatorname{Re}(\vct{h}^*\vct{a}_r\vct{a}_r^*\vct{x})|\vct{a}_r^*\vct{h}|^2+\frac{9}{10}s^2\abs{\vct{a}_r^*\vct{h}}^4\right)\geq \frac{1}{\alpha}+\frac{(1-\delta)}{4}.
\end{align}
By Corollary \ref{corineq}, with high probability,
\begin{align*} 
\frac{1}{m}\sum_{r=1}^m\Real(\vct{h}^*\vct{a}_r\vct{a}_r^*\vct{x})^2\le\frac{1+\delta}{2}+\frac{3}{2}\Real(\vct{x}^*\vct{h})^2,
\end{align*}
holds for all $\vct{h}$ obeying $\twonorm{\vct{h}}=1$. Therefore, to
establish the local curvature condition \eqref{LCC} it suffices to
show that
\begin{multline}
\label{eq:reduced_assumption1_alt}
\frac{1}{m}\sum_{r=1}^m \left(\frac{5}{2}\operatorname{Re}(\vct{h}^*\vct{a}_r\vct{a}_r^*\vct{x})^2+3s\operatorname{Re}(\vct{h}^*\vct{a}_r\vct{a}_r^*\vct{x})|\vct{a}_r^*\vct{h}|^2+\frac{9}{10}s^2|\vct{a}_r^*\vct{h}|^4\right)\ge\left(\frac{1}{\alpha}+\frac{1}{2}\right)+\frac{3}{4}\operatorname{Re}(\vct{x}^*\vct{h})^2.
\end{multline}
We will establish \eqref{eq:reduced_assumption1_alt} for different
measurement models and different values of $\epsilon$. Below, it shall
be convenient to use the shorthand
\begin{align*}
Y_r(\vct{h},s):=&\frac{5}{2}\Real(\vct{h}^*\vct{a}_r\vct{a}_r^*\vct{x})^2+3s\Real(\vct{h}^*\vct{a}_r\vct{a}_r^*\vct{x})|\vct{a}_r^*\vct{h}|^2+\frac{9}{10}s^2|\vct{a}_r^*\vct{h}|^4,\\
\avg{Y_r(\vct{h},s)}:=&\frac{1}{m}\sum_{r=1}^mY_r(\vct{h},s).
\end{align*}

\subsubsection{Proof of \eqref{eq:reduced_assumption1_alt} with
  $\epsilon={1}/{8\sqrt{n}}$ in the Gaussian and CDP models}

Set $\epsilon={1}/{8\sqrt{n}}$. We show that with high probability,
\eqref{eq:reduced_assumption1_alt} holds for all $\vct{h}$ satisfying
Im$(\vct{h}^*\vct{x})=0$, $\|\vct{h}\|_2=1$, $0\leq s \leq \epsilon$,
$\delta\le0.01$, and $\alpha\ge 30$. First, note that by
Cauchy-Schwarz inequality, 
\begin{align}
\label{LCCtemp1}
\avg{Y_r(\vct{h},s)}&\geq \frac{5}{2m} \sum_{r=1}^m \Real(\vct{h}^*\vct{a}_r\vct{a}_r^*\vct{x})^2- \frac{3s}{m} \sqrt{\sum_{r=1}^m \Real(\vct{h}^*\vct{a}_r\vct{a}_r^*\vct{x})^2} \sqrt{\sum_{r=1}^m|\vct{a}_r^*\vct{h}|^4}+\frac{9}{10}\frac{s^2}{m} \sum_{r=1}^m|\vct{a}_r^*\vct{h}|^4
\nonumber\\
&=\left(\sqrt{\frac{5}{2m}\sum_{r=1}^m \Real(\vct{h}^*\vct{a}_r\vct{a}_r^*\vct{x})^2}-s\sqrt{\frac{9}{10m}\sum_{r=1}^m|\vct{a}_r^*\vct{h}|^4}\right)^2\nonumber\\
&\ge\frac{5}{4m}\sum_{r=1}^m \Real(\vct{h}^*\vct{a}_r\vct{a}_r^*\vct{x})^2-\frac{9s^2}{10m}\sum_{r=1}^m|\vct{a}_r^*\vct{h}|^4.
\end{align}
The last inequality follows from $(a-b)^2\ge\frac{a^2}{2}-b^2$. By
Corollary \ref{corineq}, 
\begin{align}
\label{LCCinter1}
\frac{1}{m}\sum_{r=1}^m \Real(\vct{h}^*\vct{a}_r\vct{a}_r^*\vct{x})^2\ge\frac{1-\delta}{2}+\frac{3}{2}\Real(\vct{x}^*\vct{h})^2
\end{align}
holds with high probability for all $\vct{h}$ obeying
$\twonorm{\vct{h}}=1$. Furthermore, by applying Lemma
\ref{concenCovariance}, 
\begin{align}
\label{LCCinter2}
\frac{1}{m}\sum_{r=1}^m|\vct{a}_r^*\vct{h}|^4\le(\max_r\twonorm{\vct{a}_r}^2)\left(\frac{1}{m}\sum_{r=1}^m|\vct{a}_r^*\vct{h}|^2\right)\le6(1+\delta)n
\end{align}
holds with high probability. Plugging \eqref{LCCinter1} and
\eqref{LCCinter2} in \eqref{LCCtemp1} yields
\begin{align*}
  \avg{Y_r(\vct{h},s)}&\ge\frac{15}{8}\Real(\vct{x}^*\vct{h})^2+\frac{5}{8}(1-\delta)-\frac{27}{5}s^2(1+\delta)n.
\end{align*}
\eqref{eq:reduced_assumption1_alt} follows by using $\alpha\ge30$, $\epsilon=\frac{1}{8\sqrt{n}}$ and $\delta=0.01$.

\subsubsection{Proof of \eqref{eq:reduced_assumption1_alt} with
  $\epsilon={1}/{8}$ in the Gaussian model}

Set $\epsilon={1}/{8}$. We show that with high probability,
\eqref{eq:reduced_assumption1_alt} holds for all $\vct{h}$ satisfying
$\Imag(\vct{h}^*\vct{x})=0$, $\|\vct{h}\|_2=1$, $0\leq s \leq
\epsilon$, $\delta\le2$, and $\alpha\ge 8$. To this end, we first
state a result about the tail of a sum of i.i.d.~random
variables. Below, $\Phi$ is the cumulative distribution function of a
standard normal variable. 
\begin{lemma}[\cite{Bentkus2003}]\label{teo:one-sided}Suppose
  $X_1,X_2,\ldots,X_m$ are i.i.d.~real-valued random variables obeying
  $X_r\leq b$ for some nonrandom $b>0$, $\E X_r=0$, and $\E
  X_r^2=v^2$. Setting $\sigma^2=m\max(b^2, v^2)$, 
\[
\P(X_1+\ldots+X_m\geq y)\leq
\min\left(\exp\left(-\frac{y^2}{2\sigma^2}\right),\, 
  c_0\left(1-\Phi({y}/{\sigma})\right)\right)
\]
where one can take $c_0= 25$.
\end{lemma}
To establish \eqref{eq:reduced_assumption1_alt} we first prove it for
a fixed $\vct{h}$, and then use a covering argument. Observe that 
\[
Y_r:=Y_r(\vct{h},s)=\left(\sqrt{\frac{5}{2}}\Real(\vct{h}^*\vct{a}_r\vct{a}_r^*\vct{x})+\sqrt{\frac{9}{10}}s\abs{\vct{a}_r^*\vct{h}}^2\right)^2.
\]
By Lemma \ref{usefulexpG},
\[
\E[\Real(\vct{h}^*\vct{a}_r\vct{a}_r^*\vct{x})^2]=\frac{1}{2}+\frac{3}{2}(\Real(\vct{x}^*\vct{h}))^2 \text{ and } \E[\Real(\vct{h}^*\vct{a}_r\vct{a}_r^*\vct{x})|\vct{a}_r^*\vct{h}|^2]=2\Real(\vct{u}^*\vct{v}),
\]
compare \eqref{term1} and \eqref{term2}.  Therefore, using
$s\le\frac{1}{8}$,
\[
\mu_r=\E Y_r=\frac{5}{4}(1+3\Real(\vct{x}^*\vct{h})^2)+6s\Real(\vct{x}^*\vct{h})+\frac{27}{10}s^2<6.
\]
Now define $X_r=\mu_r-Y_r$. First, since $Y_r\geq 0$,
$X_r\le\mu_r<6$. Second, we bound $\E X_r^2$ using Lemma
\ref{usefulexpG} and Holder's inequality with $s\le{1}/{8}$:
\begin{align*}
  \E X_r^2 \le \E Y_r^2 & =
  \frac{25}{4}\E[\Real(\vct{h}^*\vct{a}_r\vct{a}_r^*\vct{x})^4]+\frac{81}{100}s^4\E[|\vct{a}_r^*\vct{h}|^8]+\frac{27}{2}s^2\E[\Real(\vct{h}^*\vct{a}_r\vct{a}_r^*\vct{x})^2\abs{\vct{a}_r^*\vct{h}}^4]\\
  &\qquad \qquad + 15s\E[\Real(\vct{h}^*\vct{a}_r\vct{a}_r^*\vct{x})^3|\vct{a}_r^*\vct{h}|^2]+\frac{27}{5}s^3\E[\Real(\vct{h}^*\vct{a}_r\vct{a}_r^*\vct{x})\abs{\vct{a}_r^*\vct{h}}^6]\\
  & \le \frac{25}{4}\sqrt{\E[\abs{\vct{a}_r^*\vct{h}}^8]\E[\abs{\vct{a}_r^*\vct{x}}^8]}+\frac{81}{100}s^4\E[|\vct{a}_r^*\vct{h}|^8]+\frac{27}{2}s^2\sqrt{\E[\abs{\vct{a}_r^*\vct{h}}^{12}]\E[\abs{\vct{a}_r^*\vct{x}}^4]}\\
  &\qquad \qquad +15s\sqrt{\E[\abs{\vct{a}_r^*\vct{h}}^{10}]\E[\abs{\vct{a}_r^*\vct{x}}^6]}+\frac{27}{5}s^3\sqrt{\E[\abs{\vct{a}_r^*\vct{h}}^{14}]\E[\abs{\vct{a}_r^*\vct{x}}^2]}\\
  & < 20s^4+543s^3+513s^2+403s+150\\
  & < 210.
\end{align*}
We have all the elements to apply Lemma \ref{teo:one-sided} with
$\sigma^2=m\max(9^2,210)=210m$ and $y={m}/{4}$:
\begin{align*}
\mathbb{P}\big(m\mu-\sum_{r=1}^mY_r\ge\frac{m}{4}\big)\le e^{-3\gamma m}
\end{align*}
with $\gamma=1/1260$. Therefore, with probability at least
$1-e^{-3\gamma m}$, we have
\begin{align}
\label{bigtemp}
\avg{\mtx{Y}_r(\vct{h},s)}\ge& \frac{5}{4}(1+3\Real(\vct{x}^*\vct{h})^2)+6s\Real(\vct{x}^*\vct{h})+2.7s^2-\frac{1}{4}\nonumber\\
\ge&\frac{3}{4}+\frac{1}{16}+\frac{3}{4}\Real(\vct{x}^*\vct{h})^2+3\left(\Real(\vct{x}^*\vct{h})+s\right)^2+\left(\frac{3}{16}-\frac{3}{10}s^2\right)\nonumber\\
\ge&\frac{3}{4}+\frac{1}{16}+\frac{3}{4}\Real(\vct{x}^*\vct{h})^2.
\end{align}
provided that $s\le1/8$.  The inequality above holds for a
fixed vector $\vct{h}$ and a fixed value $s$. To prove \eqref{eq:reduced_assumption1_alt}
for all $s\le 1/8$ and all $\vct{h}\in\C^n$ with $\twonorm{\vct{h}}=1$, define
\begin{align*}
p_r(\vct{h},s):=&\sqrt{\frac{5}{2}}\Real(\vct{h}^*\vct{a}_r\vct{a}_r^*\vct{x})+\sqrt{\frac{9}{10}}s\abs{\vct{a}_r^*\vct{h}}^2.
\end{align*}
Using the fact that $\max_r\twonorm{\vct{a}_r}\le\sqrt{6n}$ and $s\le
{1}/{8}$, we have
$\abs{p_r(\vct{h},s)}\le2\abs{\vct{a}_r^*\vct{h}}\abs{\vct{a}_r^*\vct{x}}+s\abs{\vct{a}_r^*\vct{h}}^2\le13n$. Moreover,
for any $\vct{u},\vct{v}\in\C^n$ obeying
$\twonorm{\vct{u}}=\twonorm{\vct{v}}=1$,
\begin{align*}
\abs{p_r(\vct{u},s)-p_r(\vct{v},s)}\le\abs{\sqrt{\frac{5}{2}}\Real\left((\vct{u}-\vct{v})^*\vct{a}_r\vct{a}_r^*\vct{x}\right)}+\sqrt{\frac{9}{10}}s\left(\abs{\vct{a}_r^*\vct{u}}+\abs{\vct{a}_r^*\vct{v}}\right)\abs{\vct{a}_r^*(\vct{u}-\vct{v})}\le\frac{27}{2}n\twonorm{\vct{u}-\vct{v}}.
\end{align*}
Introduce
\begin{align*}
q(\vct{h},s):=\frac{1}{m}\sum_{r=1}^mp_r(\vct{h},s)^2-\frac{3}{4}\Real(\vct{x}^*\vct{h})^2=\avg{Y_r(\vct{h},s)}-\frac{3}{4}\Real(\vct{x}^*\vct{h})^2.
\end{align*}
For any $\vct{u},\vct{v}\in\C^n$ obeying
$\twonorm{\vct{u}}=\twonorm{\vct{v}}=1$,
\begin{align}
\label{percover1}
\abs{q(\vct{u},s)-q(\vct{v},s)} & = \abs{\frac{1}{m}\sum_{r=1}^m
  (p_r(\vct{u},s)-p_r(\vct{v},s))(p_r(\vct{u},s)+p_r(\vct{v},s))-\frac{3}{4}\Real(\vct{x}^*(\vct{u}-\vct{v}))Re(\vct{x}^*(\vct{u}
  + \vct{v}))}
\nonumber\\
& \le \frac{27 n}{2} \times 2 \times13n \twonorm{\vct{u}-\vct{v}} +
\frac{3}{2}\twonorm{\vct{u}-\vct{v}} \nonumber\\ & = \left(351n^2+\frac{3}{2}\right)\twonorm{\vct{u}-\vct{v}}.
\end{align}
Therefore, for any $\vct{u},\vct{v}\in\C^n$ obeying
$\twonorm{\vct{u}}=\twonorm{\vct{v}}=1$ and
$\twonorm{\vct{u}-\vct{v}}\le\eta:=\frac{1}{6000 n^2}$, we have
\begin{align}
\label{percov}
q(\vct{v},s)\ge q(\vct{u},s)-\frac{1}{16}.
\end{align}
Let $\mathcal{N}_\eta$ be an $\eta$-net for the unit sphere of $\C^n$
with cardinality obeying $|\mathcal{N}_\eta|\le
(1+\frac{2}{\eta})^{2n}$. Applying \eqref{bigtemp} together with the
union bound we conclude that for all $\vct{u}\in\mathcal{N}_\eta$ and a fixed $s$
\begin{align}
\label{unioncov}
\mathbb{P}\left(q(\vct{u},s)\ge\frac{3}{4}+\frac{1}{16}\right) & \ge 1-|\mathcal{N}_\eta|e^{-3\gamma m}\nonumber\\
& \ge 1-(1+12000n^2)^ne^{-3\gamma m}\nonumber\\
& \ge 1-e^{-2\gamma m}.
\end{align}
The last line follows by choosing $m$ such that $m\ge c\cdot n\log n$,
where $c$ is a sufficiently large constant. Now for any $\vct{h}$ on
the unit sphere of $\C^n$, there exists a vector
$\vct{u}\in\mathcal{N}_\eta$ such that
$\twonorm{\vct{h}-\vct{u}}\le\eta$. By combining \eqref{percov} and
\eqref{unioncov}, $q(\vct{h},s)\ge\frac{3}{4}$ holds with probability at least $1-e^{-2\gamma m}$ for all $\vct{h}$ with unit Euclidean norm and for a fixed $s$. Applying a similar covering number argument over $s\le1/8$ we can further conclude that for all $\vct{h}$ and $s$
\begin{align*}
q(\vct{h},s)\ge&\frac{3}{4}+\frac{1}{16}-\frac{1}{16}>\frac{5}{8}\quad\Rightarrow\quad\avg{Y_r(\vct{h},s)}\ge \left(\frac{1}{8}+\frac{1}{2}\right)+\frac{3}{4}\Real(\vct{x}^*\vct{h})^2,
\end{align*}
holds with probability at least $1-e^{\gamma m}$ as long as $m\ge c\cdot n \log n$. This concludes the
proof of \eqref{eq:reduced_assumption1_alt} with $\alpha\ge 8$.

\subsection{Proof of the local smoothness condition}
For any $\vct{z}\in E(\epsilon)$, we want to prove \eqref{LSC}, which
is equivalent to proving that for all $\vct{u}\in\C^n$ obeying
$\twonorm{\vct{u}}=1$, we have
\[
\left|\left(\nabla f(\vct{z})\right)^*\vct{u}\right|^2\leq \beta\left(\frac{(1-\delta)}{4}\text{dist}^2(\vct{z},\vct{x})+\frac{1}{10m}\sum_{r=1}^m\abs{\vct{a}_r^*(\vct{z}-e^{i\phi(\vct{z})}\vct{x})}^4\right).
\]
Recall that 
\begin{align*}
\nabla f(\vct{z})=\frac{1}{m}\sum_{r=1}^m\left(\abs{\langle\vct{a}_r,\vct{z}\rangle}^2-y_r\right)(\vct{a}_r\vct{a}_r^*)\vct{z}
\end{align*} 
and define 
\begin{align*}
g(\vct{h},\vct{w},s)=\frac{1}{m}\sum_{r=1}^m \bigg(&2\Real(\vct{h}^*\vct{a}_r\vct{a}_r^*\vct{x})\Real(\vct{w}^*\vct{a}_r\vct{a}_r^*\vct{x})+s|\vct{a}_r^*\vct{h}|^2\Real(\vct{w}^*\vct{a}_r\vct{a}_r^*\vct{x}) \nonumber
\\
&+2s\Real(\vct{h}^*\vct{a}_r\vct{a}_r^*\vct{x})\Real(\vct{w}^*\vct{a}_r\vct{a}_r^*\vct{h})+s^2|\vct{a}_r^*\vct{h}|^2\Real(\vct{w}^*\vct{a}_r\vct{a}_r^*\vct{h})\bigg).
\end{align*}
Define $\vct{h}:=e^{-\phi(\vct{z})}\vct{z}-\vct{x}$ and $\vct{w}:=e^{-i \phi(\vct{z})} \vct{u}$, to establish \eqref{LSC} it suffices to prove that
\begin{align}
\label{temphaa}
\abs{g(\vct{h},\vct{w},1)}^2\leq \beta\left( \frac{1-\delta}{4}\twonorm{\vct{h}}^2+\frac{1}{10m}\sum_{r=1}^m\abs{\vct{a}_r^*\vct{h}}^4\right).
\end{align}
holds for all $\vct{h}$ and $\vct{w}$ satisfying Im$(\vct{h}^*\vct{x})=0$, $\twonorm{\vct{h}}\leq \epsilon$ and $\twonorm{\vct{w}}=1$. Equivalently, we only need to prove  for all $\vct{h}$ and $\vct{w}$ satisfying Im$(\vct{h}^*\vct{x})=0$, $\twonorm{\vct{h}}=\twonorm{\vct{w}}=1$ and $\forall s: 0\leq s \leq \epsilon$,
\begin{align}
\label{LSCalt}
\abs{g(\vct{h},\vct{w},s)}^2\leq \beta\left( \frac{1-\delta}{4}+\frac{s^2}{10m}\sum_{r=1}^m\abs{\vct{a}_r^*\vct{h}}^4\right).
\end{align}
Note that since $(a+b+c)^2\le 3(a^2+b^2+c^2)$
\begin{align}
\label{intermediateLSC}
\abs{g(\vct{h},\vct{w},s)}^2&\le\Bigg|\frac{1}{m}\sum_{r=1}^m \bigg(2|\vct{h}^*\vct{a}_r||\vct{w}^*\vct{a}_r||\vct{a}_r^*\vct{x}|^2+3s |\vct{h}^*\vct{a}_r|^2|\vct{a}_r^*\vct{x}||\vct{w}^*\vct{a}_r|+s^2|\vct{a}_r^*\vct{h}|^3|\vct{w}^*\vct{a}_r|\bigg)\Bigg|^2\nonumber
\\
&\le 3\Bigg|\frac{2}{m}\sum_{r=1}^m |\vct{h}^*\vct{a}_r||\vct{w}^*\vct{a}_r||\vct{a}_r^*\vct{x}|^2\Bigg|^2+3\Bigg|\frac{3s}{m}\sum_{r=1}^m|\vct{h}^*\vct{a}_r|^2|\vct{a}_r^*\vct{x}||\vct{w}^*\vct{a}_r|\Bigg|^2+3\Bigg|\frac{s^2}{m}\sum_{r=1}^m|\vct{a}_r^*\vct{h}|^3|\vct{w}^*\vct{a}_r|\Bigg|^2\nonumber\\
& := 3(I_1 + I_2 + I_3). 
\end{align}
We now bound each of the terms on the right-hand side. For the first
term we use Cauchy-Schwarz and Corollary \ref{corfirst}, which give 
\begin{align}
\label{LSCterm1}
I_1 & \le
\left(\frac{1}{m}\sum_{r=1}^m\abs{\vct{a}_r^*\vct{x}}^2\abs{\vct{a}_r^*\vct{w}}^2\right)\left(\frac{1}{m}\sum_{r=1}^m\abs{\vct{a}_r^*\vct{x}}^2\abs{\vct{a}_r^*\vct{h}}^2\right)\le(2+\delta)^2.
\end{align}
Similarly, for the second term, we have
\begin{align}
\label{LSCterm2}
I_2 \le \left(\frac{1}{m}\sum_{r=1}^m\abs{\vct{a}_r^*\vct{h}}^4\right)\left(\frac{1}{m}\sum_{r=1}^m\abs{\vct{a}_r^*\vct{w}}^2\abs{\vct{a}_r^*\vct{x}}^2\right)\leq \frac{2+\delta}{m}\sum_{r=1}^m\abs{\vct{a}_r^*\vct{h}}^4.
\end{align}
Finally, for the third term we use the Cauchy-Schwarz inequality together with Lemma \ref{concenCovariance} (inequality) \eqref{corspec} to derive 
\begin{align}
\label{LSCterm3}
I_3
 \leq\left(\frac{1}{m}\sum_{r=1}^m\abs{\vct{a}_r^*\vct{h}}^3\max_r\twonorm{\vct{a}_r}\right)^2 & \leq
6n\left(\frac{1}{m}\sum_{r=1}^m\abs{\vct{a}_r^*\vct{h}}^3\right)^2
\nonumber\\
& \le 6n\left(\frac{1}{m}\sum_{r=1}^m\abs{\vct{a}_r^*\vct{h}}^4\right)
\left(\sum_{r=1}^m\abs{\vct{a}_r^*\vct{h}}^2\right)\nonumber\\
& \le \frac{6n(1+\delta)}{m}\sum_{r=1}^m\abs{\vct{a}_r^*\vct{h}}^4.
\end{align}
We now plug these inequalities into \eqref{intermediateLSC} and get 
\begin{align}
\label{intermediateLSC2}
\abs{g(\vct{h},\vct{w},s)}^2&\le12(2+\delta)^2+\frac{27s^2(2+\delta)}{m}\sum_{r=1}^m\abs{\vct{a}_r^*\vct{h}}^4+\frac{18 s^4 n(1+\delta)}{m}\sum_{r=1}^m\abs{\vct{a}_r^*\vct{h}}^4\nonumber\\
&\le \beta\left(\frac{1-\delta}{4}+\frac{s^2}{10m}\sum_{r=1}^m\abs{\vct{a}_r^*\vct{h}}^4\right), 
\end{align}
which completes the proof of \eqref{LSCalt} and, in turn, establishes 
the local smoothness condition in \eqref{LSC}. However, the last line
of \eqref{intermediateLSC2} holds as long as
\begin{align}
\label{betabound}
\beta\ge \max \left( 48\frac{(2+\delta)^2}{1-\delta},
  270(2+\delta)+180 \epsilon^2 n (1+\delta)\right).
\end{align}
In our theorems we use two different values of $\epsilon=\frac{1}{8\sqrt{n}}$ and $\epsilon=\frac{1}{8}$. Using $\delta\le 0.01$ in \eqref{betabound} we conclude that the local smoothness condition \eqref{intermediateLSC2} holds as long as
\[
\begin{array}{lll}
  \beta\ge 550 & \quad \text{for} & \quad\epsilon={1}/{(8\sqrt{n})},\\
  \beta\ge 3n+550 & \quad\text{for} & \quad\epsilon={1}/{8}.
\end{array}
\]

\subsection{Wirtinger flow initialization}
\label{initproofsec}

In this section, we prove that the WF initialization obeys
\eqref{gauss1init} from Theorem \ref{gauss1}.
Recall that
\[
\mtx{Y}:=\frac{1}{m}\sum_{r=1}^m
|\vct{a}_r^*\vct{x}|^2\vct{a}_r\vct{a}_r^*.
\]
and that Lemma \ref{concenHessian} gives
\begin{align*}
\|\mtx{Y}-(\vct{x}\vct{x}^*+\twonorm{\vct{x}}^2\mtx{I})\|\leq \delta:=0.001.
\end{align*}
Let $\tilde{\vct{z}}_0$ be the eigenvector corresponding to the top
eigenvalue $\lambda_0$ of $\mtx{Y}$ obeying $\twonorm{\tilde{\vct{z}}_0}=1$. It is easy to see that
\[
\left|\lambda_0-(|\tilde{\vct{z}}_0^*\vct{x}|^2+1)\right|=\left|\tilde{\vct{z}}_0^*\mtx{Y}\tilde{\vct{z}}_0-\tilde{\vct{z}}_0^*(\vct{x}\vct{x}^*+\mtx{I})\tilde{\vct{z}}_0\right|=\left|\tilde{\vct{z}}_0^*\left(\mtx{Y}-(\vct{x}\vct{x}^*+\mtx{I})\right)\tilde{\vct{z}}_0\right|\leq \|\mtx{Y}-(\vct{x}\vct{x}^*+\mtx{I})\|\leq \delta.
\]
Therefore,
\[
|\tilde{\vct{z}}_0^*\vct{x}|^2\geq \lambda_0-1 -\delta.
\]
Also, since $\lambda_0$ is the top eigenvalue of $\mtx{Y}$, and $\twonorm{\vct{x}}=1$, we have
\[
\lambda_0\geq \vct{x}^*\mtx{Y}\vct{x}= \vct{x}^*\left(\mtx{Y}-(\mtx{I}+\vct{x}\vct{x}^*)\right)\vct{x}+2\geq 2-\delta.
\]
Combining the above two inequalities together, we have
\[
|\tilde{\vct{z}}_0^*\vct{x}|^2\geq 1 -2\delta\quad\Rightarrow\quad
\text{dist}^2(\tilde{\vct{z}}_0,\vct{x})\le2-2\sqrt{1-2\delta}<\frac{1}{256}\quad\Rightarrow\quad\text{dist}(\tilde{\vct{z}}_0,\vct{x})\le\frac{1}{16}.
\]
Recall that $\vct{z}_0=\left(\sqrt{\frac{1}{m} \sum_{r=1}^m |\vct{a}_r^*\vct{x}|^2}\right)\tilde{\vct{z}}_0$. By Lemma \ref{concenCovariance}, equation \eqref{corspec}, with high probability we have 
\[
\abs{\twonorm{\vct{z}_0}^2-1}=\abs{\frac{1}{m}\sum_{r=1}^m |\vct{a}_r^*\vct{x}|^2-1}\le \frac{31}{256}\quad\Rightarrow\quad\abs{\twonorm{\vct{z}_0}-1}\le \frac{1}{16}.
\]
 Therefore, we have
\begin{equation*}
\text{dist}(\vct{z}_0,\vct{x})\le \twonorm{\vct{z}_0-\tilde{\vct{z}}_0}+\text{dist}(\tilde{\vct{z}}_0,\vct{x})=\abs{\twonorm{\vct{z}_0}-1}+\text{dist}(\tilde{\vct{z}}_0,\vct{x})\le\frac{1}{8}.
\end{equation*}

\subsection{Initialization via resampled Wirtinger Flow}
\label{QIINITsec}

In this section, we prove that that the output of Algorithm \ref{QIINITn} obeys \eqref{CDPinit}
from Theorem \ref{CDPthm}. Introduce
\begin{align*}
F(\vct{z})=\frac{1}{2}\vct{z}^*(\mtx{I}-\vct{x}\vct{x}^*)\vct{z}+\left(\twonorm{\vct{z}}^2-1\right)^2.
\end{align*}
By Lemma \ref{expGrad}, if $\vct{z} \in \C^n$ is a vector independent
from the measurements, then
\begin{align*}
  \E \nabla f(\vct{z};b) = \nabla F(\vct{z}).
\end{align*}
We prove that $F$ obeys a regularization condition in $E(1/8)$, namely,
\begin{align}
\label{avgreg}
\Real\left(\langle\nabla F(\vct{z}),\vct{z}-\vct{x}e^{i\phi(\vct{z})}\rangle\right)\ge \frac{1}{\alpha'}\text{dist}^2(\vct{z},\vct{x})+\frac{1}{\beta'}\twonorm{\nabla F(\vct{z})}^2.
\end{align}
Lemma \ref{concenGrad} implies that for a fixed vector $\vct{z}$, 
\begin{align}
\label{interinitres}
\Real \left(\left\langle\nabla f(\vct{z};b),\vct{z}-\vct{x}e^{i\phi(\vct{z})}\right\rangle\right)=&\Real\left(\left\langle\nabla F(\vct{z}),\vct{z}-\vct{x}e^{i\phi(\vct{z})}\right\rangle\right)+\Real\left(\left\langle\nabla f(\vct{z};b)-\nabla F(\vct{z}),\vct{z}-\vct{x}e^{i\phi(\vct{z})}\right\rangle\right)
\nonumber\\
\geq & \Real\left(\left\langle\nabla F(\vct{z}),\vct{z}-\vct{x}e^{i\phi(\vct{z})}\right\rangle\right)-  \twonorm{\nabla f(\vct{z};b)-\nabla F(\vct{z})} \text{dist}(\vct{z},\vct{x})
\nonumber\\
\geq & \Real\left(\left\langle\nabla F(\vct{z}),\vct{z}-\vct{x}e^{i\phi(\vct{z})}\right\rangle\right)-  \delta\text{dist}(\vct{z},\vct{x})^2\nonumber\\
\geq & \left(\frac{1}{\alpha'}-\delta\right)\text{dist}(\vct{z},\vct{x})^2+\frac{1}{\beta'}\twonorm{\nabla F(\vct{z})}^2,
\end{align}
holds with high probability. The last inequality follows from
\eqref{avgreg}. Applying Lemma \ref{concenGrad}, we also have
\[
\twonorm{\nabla F(\vct{z})}^2\geq \frac{1}{2}\twonorm{\nabla f(\vct{z};b)}^2    -   \twonorm{\nabla f(\vct{z};b)-\nabla F(\vct{z})}^2 \geq \frac{1}{2}\twonorm{\nabla f(\vct{z};b)}^2    -   \delta^2 \text{dist}(\vct{z},\vct{x})^2.
\]
Plugging the latter into \eqref{interinitres} yields
\begin{align*}
  \Real\left(\langle\nabla
    f(\vct{z};b),\vct{z}-\vct{x}e^{i\phi(\vct{z})}\rangle\right) & \ge
  \left(\frac{1}{\alpha'}-\frac{\delta^2}{\beta'}-\delta\right)\text{dist}^2(\vct{z},\vct{x})+\frac{1}{2\beta'}\twonorm{\nabla
    f(\vct{z};b)}\\
  &
  :=\frac{1}{\tilde{\alpha}}\text{dist}^2(\vct{z},\vct{x})+\frac{1}{\tilde{\beta}}\twonorm{\nabla
    f(\vct{z};b)}.
\end{align*}
Therefore, using the general convergence analysis of gradient descent
discussed in Section \ref{GCAsec} we conclude that for all
$\tilde{\mu}\le\tilde{\mu}_0:={2}/{\tilde{\beta}}$,
\begin{align*}
\text{dist}^2(\vct{u}_{b+1},\vct{x})\le\left(1-\frac{2\tilde{\mu}}{\tilde{\alpha}}\right)\text{dist}^2(\vct{u}_b,\vct{x}).
\end{align*}
Finally, 
\begin{align*}
B\ge-\frac{\log n}{\log \left(1-\frac{2\tilde{\mu}}{\tilde{\alpha}}\right)} \quad \Longrightarrow \quad 
\text{dist}(\vct{u}_{B},\vct{x})\le\left(1-\frac{2\tilde{\mu}}{\tilde{\alpha}}\right)^{\frac{B}{2}}\text{dist}(\vct{u}_0,\vct{x})\le\left(1-\frac{2\tilde{\mu}}{\tilde{\alpha}}\right)^{\frac{B}{2}}\frac{1}{8}\le\frac{1}{8\sqrt{n}}.
\end{align*}

It only remains to establish \eqref{avgreg}.  First, without loss of
generality, we can assume that $\phi(\vct{z})=0$, which implies
$\Real(\vct{z}^*\vct{x})=\abs{\vct{z}^*\vct{x}}$ and use
$\twonorm{\vct{z}-\vct{x}}$ in lieu of dist$(\vct{z},\vct{x})$. Set
$\vct{h}:=\vct{z}-\vct{x}$ so that $\Imag (\vct{x}^*\vct{h})=0$. This
implies
\begin{align*}
\nabla F(\vct{z})&=(\mtx{I}-\vct{x}\vct{x}^*)\vct{z}+2\left(\twonorm{\vct{z}}^2-1\right)\vct{z}\\
&= (\mtx{I}-\vct{x}\vct{x}^*)(\vct{x}+\vct{h})+2\left(\twonorm{\vct{x}+\vct{h}}^2-1\right)(\vct{x}+\vct{h})
\\
&=\left(\mtx{I}-\vct{x}\vct{x}^*\right)\vct{h}+2\left(2\Real (\vct{x}^*\vct{h})+\twonorm{\vct{h}}^2\right)(\vct{x}+\vct{h})
\\
&=(1+4(\vct{x}^*\vct{h})+2\twonorm{\vct{h}}^2)\vct{h}+(3(\vct{x}^*\vct{h})+2\twonorm{\vct{h}}^2)\vct{x}.
\end{align*}
Therefore,
\begin{align}
\label{inter1736}
\twonorm{\nabla F(\vct{z})}\leq 4\twonorm{\vct{h}}+6\twonorm{\vct{h}}^2+2\twonorm{\vct{h}}^3 \leq 5\twonorm{\vct{h}},
\end{align}
where the last inequality is due to $\twonorm{\vct{h}}\leq \epsilon
\leq {1}/{8}$.  Furthermore,
\begin{align}
\label{inter2736}
\Real\left(\langle\nabla F(\vct{z}),\vct{z}-\vct{x}\rangle\right) &= \Real \left(\left\langle (1+4(\vct{x}^*\vct{h})+2\twonorm{\vct{h}}^2)\vct{h}+(3(\vct{x}^*\vct{h})+2\twonorm{\vct{h}}^2)\vct{x}, \vct{h} \right\rangle \right)
\nonumber\\
& = \twonorm{\vct{h}}^2+2\twonorm{\vct{h}}^4+6\twonorm{\vct{h}}^2(\vct{x}^*\vct{h})+3\abs{\vct{x}^*\vct{h}}^2 \geq \frac{1}{4}\twonorm{\vct{h}}^2,
\end{align}
where the last inequality also holds because $\twonorm{\vct{h}}\leq
\epsilon \leq {1}/{8}$. Finally, \eqref{inter1736} and
\eqref{inter2736} imply
\[
\Real\left(\langle\nabla F(\vct{z}),\vct{z}-\vct{x}\rangle\right) \ge
\frac{1}{4}\twonorm{\vct{h}}^2\ge\frac{1}{8}\twonorm{\vct{h}}^2+\frac{1}{200}\twonorm{\nabla
  F(\vct{z})}^2:=\frac{1}{\alpha'}\twonorm{\vct{h}}^2+\frac{1}{\beta'}\twonorm{\nabla
  F(\vct{z})}^2,
\]
where $\alpha'=8$ and $\beta'=200$.
\small
\subsection*{Acknowledgements}
E. C. is partially supported by AFOSR under grant FA9550-09-1-0643 and
by gift from the Broadcom Foundation.  X.~L.~is supported by the
Wharton Dean's Fund for Post-Doctoral Research and by funding from the
National Institutes of Health. M.~S.~is supported by a Benchmark
Stanford Graduate Fellowship and AFOSR grant FA9550-09-1-0643.
M.~S.~would like to thank Andrea Montanari for helpful discussions and
for the class \cite{EE378Bweb} which inspired him to explore provable
algorithms based on non-convex schemes. He would also like to thank
Alexandre d'Aspremont, Fajwel Fogel and Filipe Maia for sharing some
useful code regarding 3D molecule reconstructions, Ben Recht for
introducing him to reference \cite{murty1987some}, and Amit Singer for
bringing the paper \cite{Altglobal} to his attention.

\bibliography{PRnonCVX}
\bibliographystyle{plain}

\normalsize

\newpage
\appendix

\section{Expectations and deviations}
\label{proofexpconc}

We provide here the proofs of our intermediate results. Throughout
this section we use $\mtx{D}\in\C^{n\times n}$ to denote a diagonal
random matrix with diagonal elements being i.i.d.~samples from an
admissible distribution $d$ (recall the definition \eqref{momentcond}
of an admissible random variable). For ease of exposition, we shall
rewrite \eqref{mainCDPmodel} in the form
\begin{align*}
  y_r = \left| \sum_{t = 0}^{n-1} x[t] \bar{d}_\ell(t) e^{-i2\pi k t/n}
  \right|^2=\abs{\vct{f}_k^*\mtx{D}_\ell^*\vct{x}}^2, \quad r = (\ell, k), \quad \begin{array}{l} 0 \le k \le n-1\\
    1 \le \ell \le L
  \end{array},
  \end{align*}
  where $\vct{f}_k^*$ is the $k$th row of the $n \times n$ DFT matrix
  and $\mtx{D}_\ell$ is a diagonal matrix with the diagonal entries
  given by $d_\ell(0),d_\ell(1),\ldots,d_\ell(n-1)$. In our model, the
  matricces $\mtx{D}_\ell$ are i.i.d.~copies of $\mtx{D}$.

\subsection{Proof of Lemma \ref{expectHessian}}

The proof for admissible coded diffraction patterns follows from
Lemmas 3.1 and 3.2 in \cite{PRCDP}. For the Gaussian model, it is a
consequence of the two lemmas below, whose proofs are ommitted.

\begin{lemma} \label{lemA1} Suppose the sequence $\{\vct{a}_r\}$
  follows the Gaussiam model. Then for any fixed vector
  $\vct{x}\in\C^n$,
\begin{align*}
\E\bigg(\frac{1}{m}\sum_{r=1}^m\abs{\vct{a}_r^*\vct{x}}^2\vct{a}_r\vct{a}_r^*\bigg)=\vct{x}\vct{x}^*+\twonorm{\vct{x}}^2\mtx{I}.
\end{align*}
\end{lemma}

\begin{lemma}
  Suppose the sequence $\{\vct{a}_r\}$ follows the Gaussiam
  model. Then for any fixed vector $\vct{x}\in\C^n$,
\begin{align*}
\E\bigg(\frac{1}{m}\sum_{r=1}^m(\vct{a}_r^*\vct{x})^2\bar{\vct{a}}_r\vct{a}_r^*\bigg)=2\vct{x}\vct{x}^T.
\end{align*}
\end{lemma}

\subsection{Proof of Lemma \ref{expGrad}}
Recall that
\begin{align*}
\nabla f(\vct{z})=&\frac{1}{m}\sum_{r=1}^m\left(\abs{\langle\vct{a}_r,\vct{z}\rangle}^2-y_r\right)(\vct{a}_r\vct{a}_r^*)\vct{z}=\frac{1}{m}\sum_{r=1}^m\left(\abs{\langle\vct{a}_r,\vct{z}\rangle}^2-\abs{\langle\vct{a}_r,\vct{x}\rangle}^2\right)(\vct{a}_r\vct{a}_r^*)\vct{z}.
\end{align*} 
Thus by applying Lemma 3.1 in \cite{PRCDP} (for the CDP model) and
Lemma \ref{lemA1} above (for the Gaussian model) we have
\begin{align*}
\E[\nabla f(\vct{z})] &  =  \frac{1}{m} \E \left[\sum_{r=1}^m (\abs{\vct{a}_r^*\vct{z}}^2\vct{a}_r\vct{a}_r^*\vct{z}-\abs{\vct{a}_r^*\vct{x}}^2\vct{a}_r\vct{a}_r^*\vct{z})\right]\\
& = (\vct{z}\vct{z}^*+\twonorm{\vct{z}}^2\mtx{I})\vct{z}-(\vct{x}\vct{x}^*+\mtx{I})\vct{z}\\ & = 2(\twonorm{\vct{z}}^2-1)\vct{z}+(\mtx{I}-\vct{x}\vct{x}^*)\vct{z}.
\end{align*}

\subsection{Proof of Lemma \ref{usefulexpG}}

Suppose $\vct{a}\in\C^n \sim
\mathcal{N}(0,\mtx{I}/2)+i\mathcal{N}(0,\mtx{I}/2)$. Since the law of
$\vct{a}$ is invariant by unitary transformation, we may just as well
take $\vct{v}=\vct{e}_1$ and
$\vct{u}=s_1e^{i\phi_1}\vct{e}_1+s_2e^{i\phi_2}\vct{e}_2$, where
$s_1,s_2$ are positive real numbers obeying $s_1^2+s_2^2=1$. We have
\begin{align*}
\E \big[\left(\Real(\vct{u}^*\vct{a}\vct{a}^*\vct{v})\right)^2\big]=& \E [\big(\Real(s_1e^{i\phi_1}\abs{a_1}^2+s_2e^{i\phi_2}a_1\bar{a}_2)\big)^2]= 2s_1^2\cos^2(\phi_1)+\frac{1}{2}s_2^2
\\
=& \frac{1}{2}+\frac{3}{2}s_1^2\cos^2(\phi_1)-\frac{1}{2}s_1^2\sin^2(\phi_1) = \frac{1}{2}+\frac{3}{2}\big(\Real(\vct{u}^*\vct{v})\big)^2-\frac{1}{2}\big(\Imag(\vct{u}^*\vct{v})\big)^2.
\end{align*}
and
\begin{align*}
\E \big[\left(\Real(\vct{u}^*\vct{a}\vct{a}^*\vct{v})\right)\abs{\vct{a}^*\vct{v}}^2\big]=& \E [\big(\Real(s_1e^{-i\phi_1}\abs{a_1}^2+s_2e^{-i\phi_2}\bar{a}_1a_2)\big)\abs{a_1}^2]= 2s_1\cos(\phi_1)= 2\Real(\vct{u}^*\vct{v}).
\end{align*}
The identity \eqref{term3} follows from standard normal moment
calculations.

\subsection{ Proof of Lemma \ref{concenHessian}}

\subsubsection{The CDP model}
Write the Hessian as
\[
\mtx{Y}:=\nabla^2 f(\vct{x})=\frac{1}{nL} \sum_{\ell = 1}^L \sum_{k = 1}^n \mtx{W}_k(\mtx{D}_\ell)
\]
where 
\[
\mtx{W}_k(\mtx{D}) := \begin{bmatrix} \mtx{D} & \mtx{0} \\ \mtx{0} & \mtx{D}^* \end{bmatrix}
\begin{bmatrix} 
\mtx{A}_k(\mtx{D})
& 
\mtx{B}_k(\mtx{D})
\\ 
\overline{\mtx{B}_k(\mtx{D})}
& 
\overline{\mtx{A}_k(\mtx{D})}
\end{bmatrix} 
\begin{bmatrix} \mtx{D}^* & \mtx{0} \\ \mtx{0} & \mtx{D} \end{bmatrix} 
\]
and
\[
\mtx{A}_k(\mtx{D})=\abs{\vct{f}_k^*\mtx{D}^*\vct{x}}^2\vct{f}_k\vct{f}_k^*,
\quad
\mtx{B}_k(\mtx{D})={(\vct{f}_k^*\mtx{D}^*\vct{x})}^2\vct{f}_k\vct{f}_k^T.
\]
It is useful to recall that
\[
\E \mtx{Y} = \begin{bmatrix}
  \mtx{I}+\vct{x}\vct{x}^* & 2\vct{x}\vct{x}^T\\
  2\bar{\vct{x}}\vct{x}^* & \mtx{I}+\bar{\vct{x}}\vct{x}^T
\end{bmatrix}.
\]
Now set
\begin{align*}
\widetilde{\mtx{Y}}=\frac{1}{nL} \sum_{\ell = 1}^L \sum_{k = 1}^n \mtx{W}_k(\mtx{D}_\ell)\mathbb{1}_{\{\abs{\vct{f}_k^*\mtx{D}_\ell^*\vct{x}}\le \sqrt{2R \log n}\}},
\end{align*}
where $R$ is a positive scalar whose value will be determined shortly,
and define the events 
\begin{align*}
E_1(R)&=\{\|\widetilde{\mtx{Y}}-\E\mtx{Y}\|\leq \epsilon\},\\
E_2(R)&=\{\widetilde{\mtx{Y}} =\mtx{Y} \},\\
E_3(R)&=\bigcap_{k,\ell}\big\{\abs{\vct{f}_k^*\mtx{D}_\ell^*\vct{x}}\le \sqrt{2R \log n}\big\},\\
E&=\{\|\mtx{Y}-\E\mtx{Y}\|\leq \epsilon\}.
\end{align*}
Note that $E_1\cap E_2 \subset E$. Also, if
$\abs{\vct{f}_k^*\mtx{D}_\ell^*\vct{x}}\le \sqrt{2R \log n}$ for all
pairs $(k,\ell)$, then $\widetilde{\mtx{Y}}=\mtx{Y}$ and thus
$E_3\subset E_2$. Putting all of this together gives
\begin{align}
\label{mainE}
\nonumber
\mathbb{P}(E^c)\leq \mathbb{P}(E_1^c\bigcup E_2^c)\le \mathbb{P}(E_1^c)+\mathbb{P}(E_2^c)& \le \mathbb{P}(E_1^c)+\mathbb{P}(E_3^c)\\
& \le \mathbb{P}(E_1^c)+\sum_{\ell=1}^{L_0}\sum_{k=1}^n
\mathbb{P}(\abs{\vct{f}_k^*\mtx{D}_\ell^*\vct{x}}>\sqrt{2R\log n}). 
\end{align}
A slight modification to Lemma 3.9 in \cite{PRCDP} gives
$\mathbb{P}(E_1^c)\le {1}/{n^3}$ provided $L\ge c(R)\log^3 n$ for a
sufficiently large numerical constant $c(R)$. Since Hoeffding's
inequality yields
$\mathbb{P}(\abs{\vct{f}_k^*\mtx{D}_\ell^*\vct{x}}>\sqrt{2R\log n})\le
2n^{-R}$, we have
\begin{align*}
  \mathbb{P}(E^c)\le n^{-3} + 2(nL) n^{-R}.
\end{align*}
Setting $R=4$ completes the proof.

\subsubsection{The Gaussian model}

By unitary invariance, we may take $\vct{x}=\vct{e}_1$.  
Letting $\vct{z}(1)$, be the first coordinate of a vector $\vct{z}$,
to prove Lemma \ref{concenHessian} for the Gaussian model it suffices
to prove the two inequalities,
\begin{align}
\label{eq:diag_concen}
\left\|\frac{1}{m}\sum_{r=1}^m \abs{\vct{a}_r(1)}^2
    \vct{a}_r\vct{a}_r^* -
  \left(\mtx{I}+\vct{e}_1\vct{e}_1^T\right)\right\|\leq
\frac{\delta}{4}
\end{align}
and
\begin{align}
\label{eq:offdiag_concen}
\left\|\frac{1}{m}\sum_{r=1}^m \overline{\vct{a}_r(1)}^2 \vct{a}_r\vct{a}_r^T -
  2\vct{e}_1\vct{e}_1^T\right\|\leq \frac{\delta}{4}.
\end{align}

For any $\epsilon > 0$, there is a constant $C > 0$ with the property
that $m \geq C \cdot n$ implies 
\[
\frac{1}{m}\sum_{r=1}^m(\abs{\vct{a}_r(1)}^2 - 1) \leq \epsilon, \quad
\frac{1}{m}\sum_{r=1}^m (\abs{\vct{a}_r(1)}^4 - 2) < \epsilon, \quad
\frac{1}{m}\sum_{r=1}^m \abs{\vct{a}_r(1)}^6 <10
\]
with probability at least $1-3n^{-2}$; this is a consquence of
Chebyshev's inequality. Moreover a union bound gives
\[
\max_{1\leq r\leq m} \abs{\vct{a}_r(1)} \leq \sqrt{10 \log m}
\]
with probability at least $1-n^{-2}$. Denote by $E_0$ the event on
which the above inequalities hold.  We show that there is another
event $E_1$ of high probability such that \eqref{eq:diag_concen} and
\eqref{eq:offdiag_concen} hold on $E_0 \cap E_1$.  Our proof strategy
is similar to that of Theorem 39 in \cite{vershyninNARMT}. To prove
\eqref{eq:diag_concen}, we will show that with high probability, for
any $\vct{y}\in\C^n$ obeying $\twonorm{\vct{y}}=1$, we have
\begin{align}
\label{lessthandelta}
\nonumber I_0(\vct{y}) & :=
\abs{\vct{y}^*\left(\frac{1}{m}\sum_{r=1}^m \abs{\vct{a}_r(1)}^2
    \vct{a}_r\vct{a}_r^* -
    \left(\mtx{I}+\vct{e}_1\vct{e}_1^T\right)\right) \vct{y}}\\ & =
\abs{\frac{1}{m} \sum_{r=1}^m
  \abs{\vct{a}_r(1)}^2\abs{\vct{a}_r^*\vct{y}}^2-(1+\abs{\vct{y}(1)}^2)}
\le\frac{\delta}{4}.
\end{align}

For this purpose, partition $\vct{y}$ in the form $\vct{y}=
(\vct{y}(1),\tilde{\vct{y}})$ with $\tilde{\vct{y}} \in\C^{n-1}$, and
decompose the inner product as
\[
\abs{\vct{a}_r^*\vct{y}}^2 = \left(\abs{\vct{a}_r(1)}^2\abs{{\vct{y}(1)}}^2+2 \Real
  \left(\tilde{\vct{a}}_r^*\tilde{\vct{y}}
    \vct{a}_r(1)\overline{\vct{y}(1)}\right)+\abs{\tilde{\vct{a}}_r^*\tilde{\vct{y}}}^2\right).
\]
This gives
\[
I_0(\vct{y}) = \abs{ \frac{1}{m}\sum_{r=1}^m
  (\abs{\vct{a}_r(1)}^4-2)\abs{{\vct{y}(1)}}^2+2\Real\left(\frac{1}{m} \sum_{r=1}^m
    \abs{\vct{a}_r(1)}^2\vct{a}_r(1)\overline{\vct{y}(1)} \, \tilde{\vct{a}}_r^*\tilde{\vct{y}}\right)+\frac{1}{m}\sum_{r=1}^m
  \abs{\vct{a}_r(1)}^2 \abs{\tilde{\vct{a}}_r^*\tilde{\vct{y}}}^2 -
  \twonorm{\tilde{\vct{y}}}^2},
\]
which follows from $\abs{\vct{y}(1)}^2+\twonorm{\tilde{\vct{y}}}^2=1$
since $\vct{y}$ has unit norm. This gives
\begin{align}
\label{fourterms}
I_0(\vct{y}) &\le \abs{ \frac{1}{m}\sum_{r=1}^m\abs{\vct{a}_r(1)}^2 -
  1}\twonorm{\tilde{\vct{y}}}^2 +\abs{
  \frac{1}{m}\sum_{r=1}^m\abs{\vct{a}_r(1)}^4 - 2}\abs{{\vct{y}(1)}}^2
\nonumber \\
& \qquad \qquad \qquad 
+2\abs{\frac{1}{m} \sum_{r=1}^m
  \abs{\vct{a}_r(1)}^2\vct{a}_r(1)\overline{\vct{y}(1)}\, \tilde{\vct{a}}_r^*\tilde{\vct{y}}}
+\abs{\frac{1}{m}\sum_{r=1}^m \abs{\vct{a}_r(1)}^2
  \left(\abs{\tilde{\vct{a}}_r^*\tilde{\vct{y}}}^2-\twonorm{\tilde{\vct{y}}}^2\right)}
\nonumber\\
&\leq 2\epsilon+2\abs{\frac{1}{m} \sum_{r=1}^m
  \abs{\vct{a}_r(1)}^2\vct{a}_r(1)\overline{\vct{y}(1)}\, \tilde{\vct{a}}_r^*\tilde{\vct{y}}}
+\abs{\frac{1}{m}\sum_{r=1}^m \abs{\vct{a}_r(1)}^2
  \left(\abs{\tilde{\vct{a}}_r^*\tilde{\vct{y}}}^2-\twonorm{\tilde{\vct{y}}}^2\right)}.
\end{align}

We now turn our attention to the last two terms of
\eqref{fourterms}. For the second term, the ordinary Hoeffding's
inequality (Proposition 10 in \cite{vershyninNARMT}) gives that for
any constants $\delta_0$ and $\gamma$, there exists a constant
$C(\delta_0, \gamma)$, such that for $m\ge C(\delta_0, \gamma)
\sqrt{n\left(\sum_{r=1}^m\abs{\vct{a}_r(1)}^6\right)}$,
\begin{align*}
\abs{\frac{1}{m} \sum_{r=1}^m (\abs{\vct{a}_r(1)}^2\vct{a}_r(1)\overline{\vct{y}(1)})\tilde{\vct{a}}_r^*\tilde{\vct{y}}}\leq \delta_0|{\vct{y}(1)}|\twonorm{\tilde{\vct{y}}} \leq \delta_0
\end{align*}
holds with probability at least $1-3e^{- 2\gamma n}$. To control the
final term, we apply the Bernstein-type inequality (Proposition 16 in
\cite{vershyninNARMT}) to assert the following: for any positive
constants $\delta_0$ and $\gamma$, there exists a constant
$C(\delta_0, \gamma)$, such that for $m\geq C(\delta_0, \gamma) (
\sqrt{n (\sum_{r=1}^m \abs{\vct{a}_r(1)}^4)}+n\max_{r=1}^m |\vct{a}_r(1)|^2)$,
\[
\abs{\frac{1}{m}\sum_{r=1}^m \abs{\vct{a}_r(1)}^2
  \left(\abs{\tilde{\vct{a}}_r^*\tilde{\vct{y}}}^2-\twonorm{\tilde{\vct{y}}}^2\right)}\leq
\delta_0 \twonorm{\tilde{\vct{y}}}^2 \leq \delta_0
\]
holds with probability at least $1-2e^{-2\gamma n}$.

Therefore, for any unit norm vector $\vct{y}$, 
\begin{align}
 \label{fixedy}
 I_0(\vct{y}) \leq 2\epsilon+ 2\delta_0
 \end{align}
 holds with probability at least $1-5e^{-2\gamma n}$.  By Lemma 5.4 in
 \cite{vershyninNARMT}, we can bound the operator norm via an
 $\epsilon$-net argument:
\begin{align*}
  \underset{\vct{y}\in\C^n}{\max} \, I_0(\vct{y}) \le 
2\, \underset{\vct{y}\in
    \mathcal{N}}{\max} \, I_0(\vct{y}) \le 4\epsilon+4\delta_0,
\end{align*}
where $\mathcal{N}$ is an $1/4$-net of the unit sphere in $\C^n$.  By
applying the union bound and choosing appropriate $\delta_0$,
$\epsilon$ and $\gamma$, \eqref{eq:diag_concen} holds with probability
at least $1-5e^{-\gamma n}$, as long as $m\geq C'(\sqrt{n \,
  \sum_{r=1}^m\abs{\vct{a}_r(1)}^6}+ \sqrt{n \sum_{r=1}^m \abs{\vct{a}_r(1)}^4} +
n\max_{1\leq r \leq m} |\vct{a}_r(1)|^2)$. On $E_0$ this inequality follows
from $m \geq C \cdot n \log n$ provided $C$ is sufficiently large.  In
conclusion, \eqref{eq:diag_concen} holds with probability at least
$1-5e^{-\gamma n}-{4}n^{-2}$.

The proof of \eqref{eq:offdiag_concen} is similar. The only difference
is that the random matrix is not Hermitian, so we work with 
\[
I_0(\vct{u},\vct{v}) = \abs{\vct{u}^* \left(\frac{1}{m}\sum_{r=1}^m
    \overline{\vct{a}_r(1)}^2 \, \vct{a}_r\vct{a}_r^T - 2\vct{e}_1\vct{e}_1^T\right)
  \vct{v}},
\]
where $\vct{u}$ and $\vct{v}$ are unit vectors.

\subsection{Proof of Corollary \ref{corineq}}

It follows from $\|\nabla^2 f(\vct{x})-\E[\nabla^2 f(\vct{x})]\|\le
\delta$ that $\nabla^2 f(\vct{x})\preceq \E[\nabla^2 f(\vct{x})]
+\delta\mtx{I}$. Therefore, using the fact that for any complex scalar
$c$, $\Real(c)^2=\frac{1}{2}\abs{c}^2+\frac{1}{2}\Real(c^2)$, we have
\begin{align*}
\frac{1}{m}\sum_{r=1}^m \Real(\vct{h}^*\vct{a}_r\vct{a}_r^*\vct{x})^2
=&\frac{1}{4}\sum_{r=1}^m\begin{bmatrix}\vct{h}\\\bar{\vct{h}}\end{bmatrix}^*\begin{bmatrix}\abs{\vct{a}_r^*\vct{x}}^2\vct{a}_r\vct{a}_r^* &(\vct{a}_r^*\vct{x})^2\vct{a}_r\vct{a}_r^T \\ (\overline{\vct{a}_r^*\vct{x}})^2\bar{\vct{a}}_r\vct{a}_r^* & \abs{\vct{a}_r^*\vct{x}}^2\bar{\vct{a}}_r\vct{a}_r^T\end{bmatrix}\begin{bmatrix}\vct{h}\\\bar{\vct{h}}\end{bmatrix}\\
\preceq& \frac{1}{4}\begin{bmatrix}\vct{h}\\\bar{\vct{h}}\end{bmatrix}^*\left(\mtx{I}_{2n}+\frac{3}{2}\begin{bmatrix}\vct{x}\\\bar{\vct{x}}\end{bmatrix}\begin{bmatrix}\vct{x}\\\bar{\vct{x}}\end{bmatrix}^*-\frac{1}{2}\begin{bmatrix}\vct{x}\\-\bar{\vct{x}}\end{bmatrix}\begin{bmatrix}\vct{x}\\-\bar{\vct{x}}\end{bmatrix}^*\right)\begin{bmatrix}\vct{h}\\\bar{\vct{h}}\end{bmatrix}+\frac{\delta}{4}\begin{bmatrix}\vct{h}\\\bar{\vct{h}}\end{bmatrix}^*\begin{bmatrix}\vct{h}\\\bar{\vct{h}}\end{bmatrix}\\
\preceq&\left(\frac{1}{2}\twonorm{\vct{h}}^2+\frac{3}{2}\Real(\vct{x}^*\vct{h})^2-\frac{1}{2}Im(\vct{x}^*\vct{h})^2\right)+\frac{\delta}{2}.
\end{align*}
The other inequality is established in a similar fashion.

\subsection{Proof of Corollary \ref{corfirst}}

In the proof of Lemma \ref{concenHessian}, we established that with
high probability, 
\begin{align*}
\|\frac{1}{m}\sum_{r=1}^m\abs{\vct{a}_r^*\vct{x}}^2\vct{a}_r\vct{a}_r^*-\left(\vct{x}\vct{x}^*+\twonorm{\vct{x}}^2\mtx{I}\right)\|\le \delta.
\end{align*}
Therefore,
\begin{align*}
\frac{1}{m}\sum_{r=1}^m\abs{\vct{a}_r^*\vct{x}}^2\vct{a}_r\vct{a}_r^*\succeq \left(\vct{x}\vct{x}^*+\twonorm{\vct{x}}^2\mtx{I}\right)-\delta\mtx{I}.
\end{align*}
This concludes the proof of one side. The other side is similar.

\subsection{Proof of Lemma \ref{concenGrad}}
Note that
\begin{align*}
\twonorm{\nabla f(\vct{z})-\E\nabla f(\vct{z})}=\max_{\vct{u}\in\C^n,\twonorm{\vct{u}}=1}\text{ }\langle\vct{u},\nabla f(\vct{z})-\E\nabla f(\vct{z})\rangle
\end{align*}
Therefore, to establish the concentration of $\nabla f(\vct{z})$
around its mean we proceed by bounding $\abs{\langle\vct{u},\nabla
  f(\vct{z})-\E\nabla f(\vct{z})\rangle}$. From Section \ref{gradHess}, 
\begin{align*}
\nabla f(\vct{z})=\frac{1}{m}\sum_{r=1}^m\left(\abs{\langle\vct{a}_r,\vct{z}\rangle}^2-y_r\right)(\vct{a}_r\vct{a}_r^*)\vct{z}.
\end{align*} 
Define $\vct{h}:=e^{-\phi(\vct{z})}\vct{z}-\vct{x}$ and $\vct{w}:=e^{-i \phi(\vct{z})} \vct{u}$, we have
\begin{multline}
\label{dotgrad1}
\langle \vct{u},\nabla f(\vct{z})\rangle=\frac{1}{m}\sum_{r=1}^m
\vct{w}^*\left(\left(\vct{a}_r^*\vct{x}\right)^2\vct{a}_r\vct{a}_r^T\right)\bar{\vct{h}}+\vct{w}^*\left(\abs{\vct{a}_r^*\vct{x}}^2\vct{a}_r\vct{a}_r^*\right)\vct{h}\\
+2\vct{w}^*\left(\abs{\vct{a}_r^*\vct{h}}^2\vct{a}_r\vct{a}_r^*\vct{h}\right)\vct{x}+\vct{w}^*\left(\left(\vct{a}_r^*\vct{h}\right)^2\vct{a}_r\vct{a}_r^T\right)\bar{\vct{x}}+\vct{w}^*\left(\abs{\vct{a}_r^*\vct{h}}^2\vct{a}_r\vct{a}_r^*\right)\vct{h}.
\end{multline}
By Lemma \ref{expGrad} we also have
\begin{multline}
\label{dotgrad2}
\vct{w}^*\E[\nabla f(\vct{z})]=\vct{w}^*\left(2\vct{x}\vct{x}^T\right)\bar{\vct{h}}+\vct{w}^*\left(\vct{x}\vct{x}^*+\twonorm{\vct{x}}^2\mtx{I}\right)\vct{h}\\
+2\vct{w}^*\left(\vct{h}\vct{h}^*+\twonorm{\vct{h}}^2\mtx{I}\right)\vct{x}+\vct{w}^*\left(2\vct{h}\bar{\vct{h}}^T\right)\bar{\vct{x}}
+\vct{w}^*\left(\vct{h}\vct{h}^*+\twonorm{\vct{h}}^2\mtx{I}\right)\vct{h}.
\end{multline}
Combining \eqref{dotgrad1} and \eqref{dotgrad2} together with the
triangular inequality and Lemma \ref{concenHessian} give 
\begin{align*}
  \abs{\langle\vct{u},\nabla f(\vct{z})-\E[\nabla f(\vct{z})]} & \le\abs{\vct{w}^*\left(\frac{1}{m}\sum_{r=1}^m\left(\vct{a}_r^*\vct{x}\right)^2\vct{a}_r\vct{a}_r^T-2\vct{x}\vct{x}^T\right)\bar{\vct{h}}}\\
  &\qquad  +\abs{\vct{w}^*\left(\frac{1}{m}\sum_{r=1}^m\abs{\vct{a}_r^*\vct{x}}^2\vct{a}_r\vct{a}_r^*-(\vct{x}\vct{x}^*+\twonorm{\vct{x}}^2\mtx{I})\right)\vct{h}}\\
  &\qquad   +2\abs{\vct{w}^*\left(\frac{1}{m}\sum_{r=1}^m\abs{\vct{a}_r^*\vct{h}}^2\vct{a}_r\vct{a}_r^*-\left(\vct{h}\vct{h}^*+\twonorm{\vct{h}}^2\mtx{I}\right)\right)\vct{x}}\\
  &\qquad  +\abs{\vct{w}^*\left(\frac{1}{m}\sum_{r=1}^m\left(\vct{a}_r^*\vct{h}\right)^2\vct{a}_r\vct{a}_r^T-2\vct{h}\bar{\vct{h}}^T\right)\bar{\vct{x}}}\\
  &\qquad  +\abs{\vct{w}^*\left(\frac{1}{m}\sum_{r=1}^m\abs{\vct{a}_r^*\vct{h}}^2\vct{a}_r\vct{a}_r^*-\left(\vct{h}\vct{h}^*+\twonorm{\vct{h}}^2\mtx{I}\right)\right)\vct{h}}.\\
  & \le \opnorm{\frac{1}{m}\sum_{r=1}^m\left(\vct{a}_r^*\vct{x}\right)^2\vct{a}_r\vct{a}_r^T-2\vct{x}\vct{x}^T}\twonorm{\vct{h}}+\opnorm{\frac{1}{m}\sum_{r=1}^m\abs{\vct{a}_r^*\vct{x}}^2\vct{a}_r\vct{a}_r^*-(\vct{x}\vct{x}^*+\twonorm{\vct{x}}^2\mtx{I})}\twonorm{\vct{h}}
  \\
  &\qquad   +2\opnorm{\frac{1}{m}\sum_{r=1}^m\abs{\vct{a}_r^*\vct{h}}^2\vct{a}_r\vct{a}_r^*-(\vct{h}\vct{h}^*+\twonorm{\vct{h}}^2\mtx{I})}+\opnorm{\frac{1}{m}\sum_{r=1}^m\left(\vct{a}_r^*\vct{h}\right)^2\vct{a}_r\vct{a}_r^T-2\vct{h}\bar{\vct{h}}^T}
  \\
  &\qquad   +\opnorm{\frac{1}{m}\sum_{r=1}^m\abs{\vct{a}_r^*\vct{h}}^2\vct{a}_r\vct{a}_r^*-\left(\vct{h}\vct{h}^*+\twonorm{\vct{h}}^2\mtx{I}\right)}\twonorm{\vct{h}}\\
  & \le 3\delta\twonorm{\vct{h}}(1+\twonorm{\vct{h}}) \\
& \le \frac{9}{2}\delta
  \twonorm{\vct{h}}.
\end{align*}

\subsection{Proof of Lemma \ref{concenCovariance}}

The result for the CDP model follows from Lemma 3.3 in
\cite{PRCDP}. For the Gaussian model, it is a consequence of standard
results, e.g.~Theorem 5.39 in \cite{vershyninNARMT}, concerning the
deviation of the sample covariance matrix from its mean.

\section{The Power Method}
\label{comp}

We use the power method (Algorithm \ref{pow}) with a random
initialization to compute the first eigenvector of $\mtx{Y} =
\mtx{A}\, \text{diag}\{\vct{y}\}\, \mtx{A}^*$.  Since, each iteration
of the power method asks to compute the matrix-vector product
\begin{align*}
  \mtx{Y}\vct{z} = \mtx{A}\, \text{diag}\{\vct{y}\} \,
  \mtx{A}^*\vct{z}, 
\end{align*}
we simply need to apply $\mtx{A}$ and $\mtx{A}^*$ to an arbitrary
vector. In the Gaussian model, this costs $2 mn$ multiplications while
in the CDP model the cost is that of $2L$ $n$-point FFTs.  We now turn
our attention to the number of iterations required to achieve a
sufficiently accurate initialization.  

\begin{algorithm}[h]
  \caption{Power Method} 
\begin{algorithmic}
  \REQUIRE{Matrix $\mtx{Y}$}
  \STATE $\vct{v}_0$ is a random vector on the unit sphere of $\C^n$
 \FOR {$\tau= 1$ \TO $T$} 
 \STATE
 $\vct{v}_\tau=\frac{\mtx{Y}\vct{v}_{\tau-1}}{\twonorm{\mtx{Y}\vct{v}_{\tau-1}}}$
  \ENDFOR
 \ENSURE{$\tilde{\vct{z}}_0=\vct{v}_{T}$}
\end{algorithmic}
\label{pow}
\end{algorithm}

Standard results from numerical linear algebra show that after $k$
iterations of the power method, the accuracy of the eigenvector is
$\mathcal{O}(\tan \theta_0 (\lambda_2/\lambda_1)^k)$, where
$\lambda_1$ and $\lambda_2$ are the top two eigenvalues of the
positive semidefinite matrix $\vct{Y}$, and $\theta_0$ is the angle
between the initial guess and the top eigenvector. Hence, we would
need on the order of $\log(n/\epsilon) \, / \,
\log(\lambda_1/\lambda_2)$ for $\epsilon$ accuracy. Under the stated
assumptions, Lemma \ref{concenHessian} bounds below the eigenvalue gap
by a numerical constant so that we can see that few iterations of the
power method would yield accurate estimates.

\end{document}